\documentclass[12pt]{article}
\usepackage{epsfig,makeidx,amsfonts,amsmath, float}
\usepackage[round]{natbib}

\def\E{\mbox{E}}

\def\N{\mbox{N}}
\def\Ga{\mbox{Ga}}

\def\Be{\mbox{Be}}

\def\I{\mbox{I}}

\def\DP{\mbox{DP}}
\def\Pn{\mbox{Pn}}

\floatstyle{boxed}
\newfloat{alg}{htbp}{lop}
\floatname{alg}{Algorithm}

\begin{document}

\title{An adaptive truncation method for inference in Bayesian nonparametric models}

\author{J.E.~Griffin\\
School of Mathematics, Statistics and Actuarial Science,\\ University of Kent, Canterbury CT2 7NF, UK}

\maketitle

\abstract{

Many exact Markov chain Monte Carlo algorithms have been developed for posterior inference in  Bayesian nonparametric models which involve infinite-dimensional priors. However, these methods are not generic and special methodology must be developed for different classes of prior or different models. Alternatively, the infinite-dimensional prior can be truncated and  standard Markov chain Monte Carlo methods used for inference. However, the error in approximating the infinite-dimensional  posterior can be hard to control for many models. This paper describes an adaptive truncation method which allows the level of the truncation to be decided by the algorithm 
and so can avoid large  errors in  approximating the posterior. A sequence of truncated priors is constructed 
which are sampled using  Markov chain Monte Carlo methods embedded in a sequential Monte Carlo algorithm. Implementational details for infinite mixture models with stick-breaking priors and normalized random measures with independent increments priors are discussed. 
The methodology is illustrated on  infinite mixture models, a semiparametric linear mixed model and a nonparametric time series model.
}

\section{Introduction}

The popularity of Bayesian nonparametric modelling has rapidly grown with computational power in a range of application 
areas such as epidemiology, biostatistics and economics. Bayesian nonparametric models have an infinite number of parameters over which a prior is placed and so allow model complexity to grow with sample size. In many models, the prior is placed on the space of probability measures.
 An overview of work in this area is given in \cite{HjHoMuWa10}. The infinite-dimensional prior does not allow the direct use of simulation-based methods for posterior inference or for the study of some properties of the distributions drawn from the prior (particularly, functionals such as the mean). The methods developed in this paper will concentrate on the former problem of posterior inference. Simulation-based methods require a finite-dimensional parameter space and  there are two main approaches to working with such a space in the literature: marginalization and truncation. 

The first approach marginalizes over the infinite-dimensional parameter and is particularly applicable to infinite mixture models such as the Dirichlet process mixture model. In these models, marginalization leads to P\'olya urn scheme (PUS) representations of the priors which can be used to define efficient Markov chain Monte Carlo (MCMC) algorithms. Algorithms based on the PUS representation of  Dirichlet process mixture models were reviewed in \cite{MacE98} and \cite{Neal00} and similar algorithms for normalized random measures mixture models were developed in \cite{FaTe13}. These methods are limited by the unavailability of a suitable PUS for some priors.

The second method is truncation in which the infinite-dimensional prior is replaced by a finite-dimensional approximation. Approaches to the approximation of the Dirichlet process were initially studied by \cite{MuTa98} who showed how the error in total variation norm between their approximation and the infinite-dimensional prior can be chosen to be smaller than any particular value. Further results and alternative truncation methods for the Dirichlet process were developed by \cite{IshZar02}. \cite{IJ01} looked at truncating the wider-class of stick-breaking (SB) priors using total variational norm to measure truncation error and described a simple block Gibbs sampler for posterior inference. More recently, slice sampling methods have been proposed for Dirichlet process \citep{Wal07, KaGrWa08}, normalized random measures with independent increments \citep{GrWa08} and $\sigma$-stable Poisson-Kingman \citep{FaWa13} priors. In slice samplers, auxiliary variables are introduced in the posterior which lead to 
finite-dimensional distributions for all full conditionals of the  Gibbs sampler and so represent a class of random truncation methods. Importantly, these methods sample exactly from the posterior distribution and so avoid truncation errors which are usually introduced by other truncation methods. However, like the marginalization methods, it is unclear how slice sampling methods can be applied generically to models with nonparametric priors.

The purpose of this paper is to develop a method for the choice of truncation level of nonparametric priors which adapts to the complexity of the data and which can be applied generically to nonparametric models. A sequence of finite-dimensional approximations to a nonparametric prior is constructed where the level of truncation is decreasing. 
A sequential Monte Carlo (SMC) approach
\citep{DMDoJa06}
 is used to sample from the corresponding sequence of posterior distributions. Some recent work in this area has emphasized the ability of these algorithms to adapt algorithmic tuning parameters to the form of the posterior distribution whilst the algorithm is running. For example, \cite{ScCh13} construct use a sequence of tempered distributions to sample from posterior distributions on high-dimensional binary spaces. The use of an SMC method allows the sequence of temperature for the tempered distribution to be decided in the algorithm.  \cite{DMDoJa12} consider inference in approximate Bayesian computation and develop an algorithm which adaptively reduce the level of approximation of the posterior. These methods share the use of the effective sample size (ESS) as a measure of discrepancy between successive distributions which can be used to adaptively tune the model parameter. Similar ideas are also described for stochastic volatility models by \cite{JaStDoTs11}. Some theoretical aspects of these types of algorithm are discussed in 
\cite{BeJaKaTh14}.

The idea of using ESS as a measure of discrepancy will play a key role in adapting the tuning parameter (the level of truncation) in the methods developed in this paper. A truncation chosen when differences in the successive discrepancies become small will typically lead to small truncation errors. Importantly, special theory (over and above the definition of a sequence of approximating processes) is not needed to implement the method.

 The paper is organized as follows. Section 2 is a review of some Bayesian nonparametric priors and truncation methods available for them. Section 3 describes the adaptive truncation algorithm for choosing the truncation level. Section 4 shows how this algorithm can be used to simulate from some popular nonparametric mixture models. Section 5 illustrates the use of these methods in infinite mixture models, and two non-standard nonparametric mixture models.


\section{Finite truncation of infinite-dimensional priors }\label{sec:trunc}

This paper will concentrate on a particular class of infinite-dimensional priors, random probability measures, which frequently arise in Bayesian nonparametric modelling. These probability measures are usually discrete with an infinite number of atoms and arise from constructions such as transformations of L\'evy processes or SB priors. In these cases,  the random probability measure $F$ can be expressed as
\begin{equation}
F=\sum_{j=1}^{\infty} p_j \delta_{\theta_j}
\label{RPM}
\end{equation}
where $p=(p_1,p_2,\dots)$ and $\theta=(\theta_1,\theta_2,\dots)$ are sequences of random variables, $p_j>0$ for $j=1,2,\dots$,  $\sum_{j=1}^{\infty} p_j=1$, and $\delta_x$ is the Dirac delta measure which places measure 1 on $x$. Usually, it is further assumed that $p$ and $\theta$ are independent {\it a priori}. This section will review the main constructions that fall within this class and truncation methods which have been proposed.

The Dirichlet process \citep{Fer73} was originally defined as a normalized gamma process so that
\[
F(B) = \frac{G(B)}{G(\Omega)}
\]
where $B$ is a measureable set, $G$ is a Gamma process, {\it i.e.} a L\'evy process with L\'evy measure $\nu(J,\theta)=M\,J^{-1}\exp\{-J\}dJ\,H(d\theta)$ where $M>0$ and $H$ is  a probability measure 
(whose density, if it exists, is $h$) with support $\Omega$ and parameters $\psi$. Alternatively, we can write
\[
F = \frac{\sum_{j=1}^{\infty} J_j \delta_{\theta_j}}{\sum_{j=1}^{\infty} J_j}
\]
where $J=(J_1,J_2,\dots)$ are the jumps of  a L\'evy process with L\'evy density $M\,J^{-1}\exp\{-J\}$  and $\theta_j\stackrel{i.i.d.}{\sim} H$.
This construction can be naturally extended to an additive, increasing stochastic process \citep[see {\it e.g}][]{Nie-BaWalk04}. A wide, tractable 
 class of such priors can be defined by assuming that the jumps $J$ arise from a suitably-defined L\'evy process, which is 
called  the class of normalized random measures with independent increments (NRMIIs) \citep{RegLijPru03}.  Examples, other than the Dirichlet process, include the normalized inverse Gaussian (NIG) process \citep{LijMePr05} and the normalized generalized gamma (NGG) process \citep{LijMePr05b, LijMePr07}. Posterior inference for the general class is described in \cite{JamesLijPr09} who assume a general formulation where $(J_1,\theta_1),(J_2,\theta_2),\dots$ follows a L\'evy process with L\'evy density $\nu(J,\theta)=h(\theta\vert J)\eta(J)$. The process is called homogeneous if $h(\theta\vert J)=h(\theta)$ and inhomogeneous otherwise. The Dirichlet process, NIG process and NGG process are all homogeneous.

Truncated versions of these normalized processes can be constructed by truncating the L\'evy process and then normalizing.
 This leads to a well-defined random probability measure. The simulation of finite-dimensional truncations of non-Gaussian L\'evy processes is an active area of research which is reviewed by \cite{conttankov}. Two truncation methods are considered in this 
paper: the Ferguson-Klass (FK) method \citep{FergKlass72} and the compound Poisson process (CPP) approximation. 
The FK
method  generates the jumps as
\[
J_j=\zeta^{-1}(t_j),\qquad j=1,2,\dots
\]
where $\zeta(x)$ is the tail mass function of $\eta$ which is defined by $\zeta(x)=\int_x^{\infty} \eta(y)\,dy$ and $t_1,t_2,\dots$ are the arrival times of a Poisson process with intensity 1. The jumps are decreasing, {\it i.e.} $J_1>J_2>J_3>\dots$. The truncated version of the process with $N$ atoms is defined to be 
\[
F^{FK}_N = \frac{\sum_{j=1}^{N} J_j \delta_{\theta_j}}{\sum_{j=1}^{N} J_j}.
\]
\cite{ALZa13} and \cite{ALZa12} provided further discussion and variations on this type of approximation for NIG and Poisson-Dirichlet processes.

An alternative truncation method uses the CPP approximation to the L\'evy process. Let $0<L<\infty$ then the jumps larger than $L$ follow a compound Poisson process 
with intensity $\eta(x)$ for $x> L$. This allows an approximation of $F$ to be defined which includes all jumps greater than $L$,
\begin{equation}
F^{CPP}_L=\frac{\sum_{j=1}^{K} J_j \delta_{\theta_j}}{\sum_{j=1}^{K} J_j}
\label{F_PP}
\end{equation}
where $K\sim\Pn\left(\int_L^{\infty}\eta(y)\,dy\right)$ and $J_1, J_2,\dots, J_K$ are independent with $J_j$ having probability density function $\frac{\eta(x)}{\int_L^{\infty}\eta(y)\,dy}$ for $x>L$. The jumps are not ordered and the number of jumps is now a random variable whose expectation increases as $L$ decreases.

The SB construction of the Dirichlet process dates from \cite{Seth} and expresses the weights in (\ref{RPM}) 
as $p_j = V_j\prod_{k<j} (1- V_k)$ where $V_j\stackrel{i.i.d.}{\sim}\Be(1,M)$ which are independent of $\theta$ with  $\theta_j\stackrel{i.i.d.}{\sim} H$.
The use of more general forms of SB prior which assume that $V_j\stackrel{ind.}{\sim}\Be(a_j,b_j)$ were popularized by  \cite{IJ01}. They gave conditions on $a=(a_1,a_2,\dots)$ and $b=(b_1,b_2,\dots)$ for the 
process to be well-defined (so that $\sum_{j=1}^{\infty} p_j=1$ almost surely). 
A nice feature of this construction is that the weights are stochastically ordered, {\it i.e.} $\E[p_1]>\E[p_2]>\E[p_3]>\dots$.
The Poisson-Dirichlet process 
\citep{PiYo97}
arises when $a_j=1-a$ and $b_j=M+aj$. \cite{FavLijPru12} derive an SB construction for the NIG process. 

A truncated version of the prior with $N$ atoms can be defined by
\begin{equation}
F^{SB}_N=\sum_{j=1}^N p_j\delta_{\theta_j}
\label{F_SB}
\end{equation}
where $p_j= V_j\prod_{k<j} (1- V_k)$ for $j=1,\dots,N$, $V_j\stackrel{ind.}{\sim}\Be(a_j,b_j)$ for $j=1,\dots,N-1$ and $V_N=1$. This truncation is well-defined since $\sum_{j=1}^N p_j=1$.  \cite{IJ01} study this truncation and show that it converges almost surely to the infinite-dimensional prior as $N\rightarrow \infty$. \cite{MuTa98} define a similar truncation, which they term $\epsilon$-Dirichlet distribution, where $N$ is chosen to be the smallest value of $W$ for which $\sum_{j=1}^{W-1}p_j>1-\epsilon$ for some pre-specified $\epsilon$ with $F$ following a Dirichlet process.
One drawback with this truncation is that the weights may no longer be stochastically ordered since $\E[p_N]>\E[p_{N-1}]$ if $b_{N-1}>a_{N-1}$. It can be shown that the weights are never stochastically ordered for a Poisson-Dirichlet process if $a>0$. An alternative method of truncation normalizes the SB prior with $N$ atoms leading to the re-normalized stick-breaking (RSB) truncation
\begin{equation}
F^{RSB}_N=\frac{\sum_{j=1}^N p_j\delta_{\theta_j}}
{\sum_{j=1}^N p_j}=\frac{\sum_{j=1}^N p_j\delta_{\theta_j}}
{1-\prod_{j=1}^N (1-V_j)}
\label{F_RSB}
\end{equation}
where $p_j = V_j\prod_{k<j} (1- V_k)$ for $j=1,\dots,N$, $V_j\stackrel{ind.}{\sim}\Be(a_j,b_j)$ for $j=1,\dots,N$. Clearly, $p_1,\dots,p_N$ are  stochastically ordered and maintain that property of  the infinite-dimensional SB prior.

An alternative form of truncation for the Dirichlet process \citep{IshZar00, Neal00}
uses finite-dimensional (FD) truncated distribution 
\[ 
F_N^{FD}=\sum_{j=1}^N \gamma_j \delta_{\theta_j}
\]
where $\gamma_j\sim\Ga(M/N, 1)$ (where $\Ga(a,b)$ represents a gamma distribution with shape $a$ and mean $a/b$) and $\theta_j\stackrel{i.i.d.}{\sim} H$. \cite{IshZar02} show that $\int g(\theta) F_N^{FD}(d\theta)\rightarrow\int g(\theta) F(d\theta)$ in distribution for any measureable function $g$ and so we can consider that this truncation converges to the Dirichlet process.

The truncation methods described so far involve the choice of $N$ for the Ferguson-Klass methods and truncated SB priors or the choice of $L$ (the smallest jump) for the CPP approximation of the NRMII process. The success of any truncation method at approximating the infinite-dimensional prior or posterior will depend critically on this choice and so considerable effect has been devoted to the choice of these truncation parameters. 

It is important to distinguish between two motivations for truncation. The first is studying the properties of the prior distribution (particularly, mean of functionals of the distribution) and the second is posterior inference using these priors. Initial work on truncation methods was motivated by the first consideration. \cite{MuTa98} demonstrated that their $\epsilon$-Dirichlet distribution can be used to 
sample Dirichlet process functionals. They showed that the truncation error can be bounded in Prohorov distance and how this result can be used to choose the truncation parameter $\epsilon$.
\cite{GeKo02} described a method for sampling  posterior functionals of the random measure. This method uses samples generated by a PUS-based sampler for the posterior of a Dirichlet process mixture and exploits ideas from sampling prior functional of Dirichlet processes. The method was extended by \cite{IJ02}.

The second motivation for the truncation method is the approximation of the infinite-dimensional posterior
 by the posterior under a truncated prior. It is important to note that a ``good'' approximation of the prior will not necessarily lead to a ``good'' approximation of the posterior distribution. 
An example of a truncation which is designed to yield a ``good'' approximation to the posterior is the truncated SB process \citep{IJ01}.
They also consider defining a suitable value $N$ to accurately approximate the inference with the infinite-dimensional prior. Their criteria is to bound the difference in the prior predictive probability of the sample under the infinite-dimensional prior and the truncated version. They show that
\[
\parallel F(y) - F_N(y)\parallel_1 < 4\left(1-\E\left[\left(\sum_{k=1}^{N-1} p_k\right)^n\right]\right)
\]
where $\parallel \cdot\parallel_1$ represents $L_1$ distance. These expectations can be calculated in the case of Poisson-Dirichlet and Dirichlet processes which allows the value of $N$ to be chosen to control the truncation error.


\section{Adaptive truncation algorithm}

The adaptive truncation algorithm will be described for a generic infinite mixture model of the form
\begin{equation}
p(y_i\vert \phi,\lambda,\kappa)=\sum p_j(\phi) k_j(y_i\vert \phi,\lambda), \quad \phi\sim\pi(\phi\vert \kappa),\quad  \lambda\sim p(\lambda),\quad \kappa\sim p(\kappa)
\label{gen_mod}
\end{equation}
where $y=(y_1,\dots,y_n)$ are a sample of observations, $k_j(y\vert \phi, \lambda)$ is a probability density function for $y$ with parameters $\phi$ and $\lambda$, $\phi$  is an  infinite-dimensional parameter, $\pi(\phi\vert \kappa)$ is a nonparametric prior and $\lambda$ and $\kappa$ are parameters of the conditional distribution of the observations and the nonparametric prior respectively. It is typical in Bayesian nonparametric methods to make inference about parameters such as $\lambda$ and $\kappa$ as well as the infinite-dimensional parameters $\phi$. The Dirichlet process mixture model would be an obvious example of a model of this form.

The adaptive truncation algorithm uses an infinite sequence of truncations of $\pi(\phi\vert \kappa)$.  The parameters 
in the first truncation are denoted $\phi_1$ and the extra parameters introduced in the $k$-th truncation will be denoted $\phi_k$. It will be helpful to define the notation $x_{j:k}=(x_j,\dots,x_k)$. The parameters in the $k$-th truncation of the nonparametric prior are  $\phi_{1:k}$, which emphasises that the parameter space is growing with the truncation level $k$.  The truncated parameters $\phi_{1:k}$ could include the first $N_k$ atoms of a SB representation or the atoms in a 
CPP approximation with   jump larger than  $L_k$. 
The prior distribution of the parameter for the $k$-th truncation will be 
 denoted by $\pi_k(\phi_{1:k}\vert \kappa)$.

I will also assume that it is relatively straightforward to sample values from $\pi_k\left(\phi_k\left\vert \phi_{1:(k-1)},\kappa\right.\right)=\left.\pi_k\left(\phi_{1:k}\left\vert \kappa\right.\right)\right/\pi_k\left(\phi_{1:(k-1)}\left\vert \kappa\right.\right)$ which is the distribution of $\phi_k$ given $\phi_{1:(k-1)}$ and $\kappa$ under the $k$-th truncation. This is  true for the truncations described in section~\ref{sec:trunc}. A sequence of models can now be constructed by replacing the infinite-dimensional prior in (\ref{gen_mod}) by the sequence of truncated priors  leading to a $k$-th model of the form
\[
p(y_i\vert \phi_{1:k},\lambda)=\sum p_j(\phi_{1:k}) k_j(y_i\vert \phi_{1:k},\lambda), \quad \phi_{1:k}\sim\pi_k(\phi_{1:k}\vert \kappa),
\]
\[
 \lambda\sim p(\lambda),\quad \kappa\sim p(\kappa)
\]
where $k_j(y\vert \phi_{1:k},\lambda)$ will typically depend on the level of truncation. 
This leads to a sequence of posterior distributions
$\pi_1(\phi_{1},\lambda,\kappa\vert y), \pi_2(\phi_{1:2},\lambda,\kappa\vert y), \dots$
for which the $k$-th posterior is
\[
\pi_k(\phi_{1:k},\lambda,\kappa\vert y)\propto \prod_{i=1}^n p(y_i\vert \phi_{1:k},\lambda)
\pi_k(\phi_{1:k}\vert \kappa)p(\lambda)p(\kappa).
\]
This sequence will  converge to the infinite-dimensional posterior (which will have the same mode of convergence as the prior).

Sequential Monte Carlo methods \citep[see][for a review]{doujoh08}
can be used to efficiently simulate
 samples from the sequence of posterior distributions.  
The steps are outlined in Algorithm 1 which uses adaptive re-sampling \citep[e.g.][]{DMDoJa06} and MCMC updating for static models \citep{chopin02}.
The parameter $b$ controls the amount of re-sampling with smaller values of $b$ implying less re-sampling. The value $b=0.7$ is chosen for the examples in this paper.
The posterior expectation of a function $f$ under the $k$-th truncated posterior will be written  $\E_{k}\left[f\left(\phi_{1:k},\lambda,\kappa\right)\right]$ and can be unbiasedly estimated  by
\begin{equation}
\frac{\sum_{j=1}^S w_k^{(j)}f\left(\phi_{1:k}^{(j)},\lambda^{(j)},\kappa^{(j)}\right)}{\sum_{j=1}^S w_k^{(j)}}
\label{SMC_approx}
\end{equation}
where $w_k^{(1)},\dots,w_k^{(S)}$ are the weights at the end of the $(k-1)$-th iteration of Algorithm 1.
 Many ways of  re-weighting the particles in step 4(a) have been described in the SMC literature.
Systematic resampling 
\citep{Kit96}
is used in this paper but other methods are described in \cite{doujoh08}.

\begin{alg}[h!]

Simulate $S$ particles, $\left(\phi_1^{(1)},\lambda^{(1)},\kappa^{(1)}\right),\dots,\left(\phi_1^{(S)},\lambda^{(S)},\kappa^{(S)}\right)$ from the posterior distribution $\pi_1(\phi_1,\lambda,\kappa\vert y)$ and set $w^{(j)}_1=1$ for $j=1,2,\dots,S$. \\
\\
At the $k$-th iteration,
\begin{enumerate}
\item Propose $\phi_{k+1}^{(j)}$ from the  transition density 
$\pi_{k+1}\left(\left.\phi_{k+1}^{(j)}\right\vert \phi_{1:k}^{(j)},\kappa^{(j)}\right)$
 for $j=1,2,\dots,S$.
\item Update the weights $w^{(1)}_{k+1},\dots,w^{(S)}_{k+1}$ according to
\[
w^{(j)}_{k+1}=w^{(j)}_k\alpha^{(j)}_k, \qquad j=1,\dots,S
\]
where
\[
\alpha^{(j)}_k=\left.\prod_{i=1}^n p\left(y_i\left\vert \phi^{(j)}_{1:(k+1)},\lambda^{(j)},\kappa^{(j)}\right.\right)
\right/\prod_{i=1}^n p\left(y_i\left\vert  \phi^{(j)}_{1:k},\lambda^{(j)},\kappa^{(j)}\right.\right)
\]
\item Calculate the effective sample size $\mbox{ESS}_k=\frac{\left(\sum_{j=1}^S w^{(j)}_{k+1}\right)^2}{\sum_{j=1}^S {w^{(j)}_{k+1}}^2}$.

\item If $\mbox{ESS}_k<bS$, 
\begin{enumerate}
\item Re-weight the particles $\left(\phi_{1:(k+1)}^{(1)},\lambda^{(1)},\kappa^{(1)}\right),\dots, \left(\phi_{1:(k+1)}^{(S)},\lambda^{(S)},\kappa^{(S)}\right)$ in proportion to the weights 
$w^{(1)}_{k+1},\dots,w^{(S)}_{k+1}$.

\item Set $w^{(j)}_{k+1}=1$ for $j=1,2,\dots,S$.

\item Update  $\left(\phi_{1:(k+1)}^{(j)},\lambda^{(j)},\kappa^{(j)}\right)$ using $m$ MCMC iterations with stationary distribution $\pi_{k+1}\left(\left.\phi_{1:(k+1)},\lambda,\kappa\right\vert y\right)$ for $j=1,\dots,S$.
\end{enumerate}

\end{enumerate}
\caption{\small The adaptive truncation algorithm}
\end{alg}

In practice, samples can only be drawn from a finite number of posteriors, {\it i.e.}
 $\pi_1(\phi_1,\lambda,\kappa\vert y), \pi_2(\phi_{1:2},\lambda,\kappa\vert y), \dots, \pi_R(\phi_{1:R},\lambda,\kappa\vert y)$. 
The truncation is made adaptive by choosing the value  of $R$ during the run of the algorithm using the output of the SMC algorithm. Intuitively, the posterior distributions $\pi_k\left(\phi_{1:k},\lambda,\kappa\vert y\right)$ will become increasingly similar as $k$ increases  since the data will tend to have less effect on the posterior of $\phi_k$ as $k$ increases.
 For example, the probability of an observation being allocated to the $k$-th cluster decreases as $k$ increases and the posterior distribution of $\phi_k$ becomes increasingly like its prior distribution. In other words,
$\pi_{k+1}\left(\phi_{1:(k+1)},\lambda,\kappa\vert y\right)$ becomes increasingly similar to  $
\tilde{\pi}_{k+1}\left(\phi_{1:(k+1)},\lambda,\kappa\right)=
\pi_k\left(\phi_{1:k},\lambda,\kappa\vert y\right)\pi_{k+1}\left(\phi_{k+1}\vert \phi_{1:k}, \kappa\right)$ as $k$ increases. 

The remaining issue is the decision of when to stop the SMC sampler. It is useful to define $\psi_{k+1}$ to be the sample of values of $\phi_{1:(k+1)}$, $\lambda$, $\kappa$ and $w_{k+1}$ at the end of the $k$-th iteration. It is assumed that a discrepancy $D(\psi_{k+1})$ between $\pi_k$ and $\pi_{k+1}$  for $j<k$ can be calculated using $\psi_k$ and $\psi_{k+1}$. The discrepancy should be positive with $D(\psi_{k+1})=0$ if $\pi_k$ and $\pi_{k+1}$ are the same
and increasing as $\pi_{k}$ and $\pi_{k+1}$  become increasingly different.
Specific examples of such discrepancies will be discussed at the end of this section.
I define the  stopping point $R$ to be the smallest $T$ for which
\begin{equation}
 D\left(\psi_{k+1}\right)<\delta \mbox{ for }k=T-m+1,\dots,T
\label{desc}
\end{equation}
  where $\delta$ and $m$ are chosen by the user. The parameter $\delta$ is positive and taken to be small (and whose value is considered in section 5) and $m$ is usually fairly small (the choice $m=3$ worked well in the examples).
This makes operational the idea of the sequence posteriors ``settling down''. It seems sensible to assume that this ``settling down'' indicates that $\pi_R(\cdot\vert y)$ is close to the posterior distribution for the infinite dimensional model.

One useful  measure of the discrepancy between these two distributions available from the SMC sampler is 
the effective sample size (ESS) \citep{liu01}, which is further investigated in this paper. This  is defined as
\[
\mbox{ESS}_{k+1}=\frac{\left(\sum_{j=1}^S w^{(j)}_{k+1}\right)^2}{\sum_{j=1}^S \left(w^{(j)}_{k+1}\right)^2}.
\]
It can be interpreted as the number of independent samples needed to produce a Monte Carlo estimate with the same accuracy as the approximation in (\ref{SMC_approx}). A
larger value of $\mbox{ESS}_k$ indicates a lower discrepancy with $\mbox{ESS}_k=S$ if the two distributions are the same.
The measure of discrepancy in (\ref{desc}) is defined, in this case, to be $D(\psi_{k+1})=\vert\mbox{ESS}_{k+1}-\mbox{ESS}_k\vert$ and we define $\delta=\epsilon S$ for a small value of $\epsilon$. The use of ESS as an algorithmic indicator is discussed in \cite{chopin02} who suggests an alternative form of ESS for SMC algorithms with MCMC steps.
Usefully, this measure relates to the sequence of posteriors of all parameters $\phi_{1:k}$, $\lambda$ and $\kappa$  rather than just the the truncated parameter $\phi_{1:k}$ and so effective inference (without large truncation error) can be made over hyperparameters as well as the nonparametric component of the model.

Alternative measures of discrepancy could be defined by looking at specific summaries of the posterior distribution. For example,  the predictive distribution of $y_{n+1}$ at a specific point $y^{\star}$. Suppose that $p_k(\cdot)$ is the predictive distribution of $y_{n+1}$ calculated using $\psi_k$ then a suitable discrepancy would be
$D(\psi_{k+1})=\vert p_{k+1}(y^{\star}) - p_k(y^{\star})\vert$. This idea could be extended to other summaries such as posterior means or variance or to weighted sums of many absolute differences between summaries.

It is important to note that this method uses the samples from $\pi_k(\cdot\vert y)$ to estimate the differences between the consecutive distributions. This will work well if the samples are representative but may lead an inappropriately small value of $R$ if the samples are not representative. The representativeness of the sample depends on a sufficiently long run of the MCMC sampler for $\pi_1$ and avoidance of very small ESS for successive steps of the algorithm. The representativeness of the sample from $\pi_1(\cdot\vert y)$ can be checked using standard methods for MCMC samplers and the ESS will not become very small if the number of atoms in the initial truncation are chosen to be sufficiently large.

'





\section{Adaptive truncation algorithms for mixture models with some specific priors}
\label{alg_examples}

This section describes adaptive truncation methods for mixture models in (\ref{gen_mod})
with specific forms of nonparametric prior.
MCMC methods typically introduce latent allocation variables $s_1,\dots,s_n$ and re-express the model as
\[
y_i\vert s_i\stackrel{ind}{\sim} k_{s_i}(y_i\vert \phi, \lambda),\qquad p(s_i=j)=p_j(\phi)\qquad i=1,\dots,n,\quad j=1,2,\dots.
\]
This approach will be used in the MCMC samplers described in this section. 

\subsection{SB priors}

\subsubsection{RSB truncation}
\label{sec:RSB}

The adaptive truncation algorithm can be used with  the RSB truncation of the SB prior by first choosing an initial truncation 
with $N_1$ atoms, {\it i.e.} $F^{RSB}_{N_1}$. This leads to an initial set of parameters $\phi_1 = (V_1,\dots,V_{N_1},\theta_1,\dots,\theta_{N_1})$. The $k$-th truncated prior in the algorithm is $F^{RSB}_{N_k}$ in (\ref{F_RSB}) where $N_k=N_1+k-1$. The extra parameters introduced in the $k$-th truncation are  $\phi_k=(V_{N_k},\theta_{N_k})$ for $k>1$
and their prior distribution is 
\[
\pi_k\left(\phi_k\left\vert \phi_{1:(k-1)}, \kappa\right.\right)=\Be\left(V_{N_k}\left\vert a_{N_k},b_{N_k}\right.\right) h\left(\theta_{N_k}\vert \psi\right)
\]
where $\kappa=(a,b,\psi)$. 
The adaptive truncation algorithm in Algorithm 1 can now be run. 
The MCMC sampler for the $k$-th truncation  introduces the latent variables $s_1,\dots,s_n$ described at the start of this section which leads to 
a joint prior distribution of $s$  given by
\[
p(s\vert V)=\prod_{i=1}^n \frac{p_{s_i}}{1-\prod_{j=1}^{N_k} (1-V_j)}.
\]
The normalization constant in the denominator  leads to non-standard full conditional distribution for $V_1,\dots,V_{N_k}$. 
The identity $\sum_{i=0}^{\infty} d^i=\frac{1}{1-d}$ if $d<1$ leads to a representation of the prior, which introduces 
latent variables $z_1,\dots,z_n$, and has density
\[
p(s,z\vert V)=\prod_{i=1}^n p_{s_i}\left(\prod_{j=1}^{N_k} (1-V_j)\right)^{z_i}
\]
where $z_i\geq 0$ for $i=1,\dots,n$ \citep[see {\it e.g.}][]{AntWal12}. It follows that $\sum p(s,z)=p(s)$ where the sum is taken over all possible values of $z$. The augmented posterior $\pi_k(\phi_k,\lambda,\kappa,s,z\vert y)$ has full
conditional distributions of standard form which allows a Gibbs sampler to be run. 
\noindent The steps of the MCMC algorithm  for $\pi_k(\phi_{1:k},\lambda,\kappa, s,z\vert y)$  are as follows.\\
\\
\noindent\textbf{Updating $s$}\\
The full conditional distribution of $s_i$ is
\[
p(s_i=j)\propto p_j\,k_j(y_i\vert \theta),\qquad j=1,2,\dots,N_k.
\]

\noindent\textbf{Updating $z$}\\
The full conditional density of $z_i$ is proportional to
\[
 \left(\prod_{j=1}^{N_k} (1-V_j)\right)^{z_i},\qquad z_i=0,1,2,\dots
\]
which is a geometric distribution with success probability $1-\prod_{j=1}^{N_k} (1-V_j)$.\\
\\
\noindent\textbf{Updating $V$}\\
The full conditional distribution of $V_j$ is $\Be\left(a_j^{\star}, b_j^{\star}\right)$ where $a_j^{\star}=a_j+\sum_{i=1}^n \I(s_i=j)$, and $b_j^{\star}=b_j+\sum_{i=1}^n \I(s_i>j) + \sum_{i=1}^n z_i$ for $j=1,2,\dots,N_k$.\\
\\
\noindent\textbf{Updating $\theta$}\\
The full conditional density of $\theta_j$ is proportional to
\[
h(\theta_j\vert \psi)\prod_{\{i\vert s_i=j\}} k_j(y_i\vert \theta).
\]

\subsubsection{SB truncation}

In a similar way to the the RSB truncation of the SB prior, an initial truncation of the infinite sum with $N_1$ atoms is chosen, 
{\it i.e.} $F^{SB}_{N_1}$,  which has initial parameters $\phi_1 = (V_1,\dots,V_{N_1-1},\theta_1,\dots,\theta_{N_1})$. The $k$-th truncated prior  is $F^{SB}_{N_k}$ in (\ref{F_SB}) where, again, $N_k=N_1+k-1$. The extra parameters introduced in the $k$-th truncation are  $\phi_k=(V_{N_k-1},\theta_{N_k})$ for $k>1$
and their prior distribution is 
\[
\pi_k\left(\phi_k\left\vert \phi_{1:(k-1)},\kappa\right.\right)=\Be\left(V_{N_k-1}\left\vert a_{N_k-1},b_{N_k-1}\right.\right) h\left(\left.\theta_{N_k}\right\vert \psi\right)
\]
where $\kappa=(a,b,\psi)$. 
 The MCMC sampler for the $k$-th truncation  is the same as the one described in the previous subsection with the exception that the 
full conditional distribution of $V_j$ is $\Be\left(a^{\star}_j, b^{\star}_j\right)$ where $a^{\star}_j=a_j+\sum_{i=1}^n \I(s_i=j)$ and $b^{\star}_j= b_j+\sum_{i=1}^n \I(s_i>j) $ for $j=1,2,\dots,N_k-1$.
This leads to a Gibbs sampler which has the exact form of  the blocked Gibbs sampler \cite{IJ01} for a truncation value $N_k$.

\subsection{L\'evy process-based models}


\subsubsection{CPP truncation}

In the CPP truncation, a sequence of truncation points $L_1,L_2,\dots$ is selected. There are many ways to choose these points. 
For example, an increment size $\xi$ could be chosen and the sequence generated using $L_k=L_{k-1}\exp\{-\xi\}=L_1\exp\{-(k-1)\xi\}$. Alternatively, a sequence which satisfies $\zeta(L_k)=\zeta(L_{k-1})+1$ would imply that, on average, one atom is added to the truncation at each iteration. The initial truncation is then $F_{L_1}^{CPP}$ which has initial parameters $\phi_1=(J_1,\dots,J_{K_{L_1}},\theta_1,\dots,\theta_{K_{L_1}})$.
The $k$-th truncated prior is $F^{CPP}_{L_k}$ in 
(\ref{F_PP}) and so the extra parameters introduced in the $k$-th truncation are  
\[
\phi_k=(J_{K_{L_{k-1}}+1},\dots,J_{K_{L_k}},\theta_{K_{L_{k-1}}+1},\dots,\theta_{K_{L_k}})
\] 
which are the jumps whose jump size is between $L_{k-1}$ and $L_k$. The distribution of $\phi_k$ conditional on $\phi_{1:(k-1)}$ and $\kappa$ under $\pi_k$ is a marked Poisson process with intensity $\eta(x)\exp\{-v x\}$ on $(L_k,L_{k-1})$ (for the jumps $J_j$) and mark distribution $H$ (for the locations $\theta_ j$).  The
MCMC samplers uses a similar approach to the slice sampler of \cite{GrWa08} by  introducing allocation variables $s_1,\dots,s_n$ and a latent variable $v$ in an augmented prior
\[
p(s_1,\dots,s_n,v)= v^{n-1}\prod_{i=1}^n J_{s_i}\exp\left\{-v\sum_{j=1}^{K_{L_k}} J_j\right\}.
\]
This  leads to the correct marginal distribution $p(s_1,\dots,s_n) = \prod_{i=1}^n \frac{J_{s_i}}{\sum_{j=1}^{K_{L_k}} J_j}$.

\noindent The steps of the MCMC algorithm for $\pi_k(\phi_k,\lambda,\kappa,s,v\vert y)$ are as follows.\\
\\
\noindent\textbf{Updating $J$ and $\theta$}\\
The parameters $J$ and $\theta$ are divided into two parts: the jumps to which observations have been allocated which will be denoted 
$J^{\star}$ and $\theta^{\star}$ and 
the jumps to which no observation has been allocated $J^{\dagger}$ and $\theta^{\dagger}$. The full conditional density of $J^{\star}_j$ is proportional to
\[
\eta\left(J_j^{\star}\right){J_j^{\star}}^{m_j}\exp\left\{-v J^{\star}_j\right\}, \qquad J^{\star}_j>L_k
\]
where $m_l$ is the number of observations allocated to the $l$-th jump
and
the full conditional density of $\theta^{\star}_j$ is proportional to
\[
h\left(\theta^{\star}_j\right)\prod_{\{i\vert s_i=j\}} k_j\left(y_i\left\vert \theta^{\star}_j\right.\right).
\]
The full conditional  of $J^{\dagger}$ is a marked Poisson process with intensity $\exp\{-v x\}\nu(x)$ on $\left(L_k,\infty\right)$ and mark distribution $H$. The sampled values are then $J=(J^{\star},J^{\dagger})$ and $\theta=(\theta^{\star},\theta^{\dagger})$.\\
\\
\noindent\textbf{Updating $s$}\\
The full conditional distribution of $s_i$ is
\[
p(s_i=j)\propto J_j\,k_j(y_i\vert \theta),\qquad j=1,2,\dots,K_{L_k}.
\]
\\
\noindent\textbf{Updating $v$}\\
The full conditional distribution of $v$ is $\Ga\left(n, \sum_{j=1}^{L_k} J_j\right)$.

\section{Examples}

\subsection{Mixture models}

The Dirichlet process mixture model \citep{Lo84}  is the most widely used Bayesian nonparametric model with many MCMC algorithms having been proposed \citep[see][for reviews]{MacE98, GriHol10}. 
Therefore, these models represent a natural benchmark for new computational methods for Bayesian nonparametric inference.
The adaptive truncation algorithm is not expected to outperform current MCMC methods for these models (in fact, its main purpose is to define a generic method for inference in non-standard nonparametric models) but it is useful to 
look at its performance in this standard model.
 The infinite  mixture models considered in this subsection used $k_j(y_i\vert \phi)=\N(y_i\vert \mu_j,\sigma^2_j)$ where $\N(x\vert \mu,\sigma^2)$ represents the density of a normally distributed  random variable with mean $\mu$ and variance $\sigma^2$. Initially, a Dirichlet process with mass parameter $M$ and centring measure $H$  with density 
$h(\mu,\sigma^{-2})=\N(\mu\vert \mu_0, \sigma^2)\,\Ga(\sigma^{-2}\vert \alpha,\beta)$ was considered
for the nonparametric prior. 
The mixture models were applied to the galaxy data, which was first introduced into the Bayesian nonparametric literature by \cite{EW95}. The observations were divided by 10\ 000.
The hyperparameters were chosen to be $\mu_0=\bar{y}$, $\sigma^2=10$, $\alpha=3$ and $\beta=0.1(\alpha-1)\hat\sigma^2$ where $\bar{y}$ and $\hat\sigma^2$ are the sample mean and  the sample variance of the observations, which were chosen for the purposes of  illustration.

\begin{figure}[h!]
\begin{center}
\includegraphics[trim=10mm 5mm 100mm 175mm, clip, scale=1]{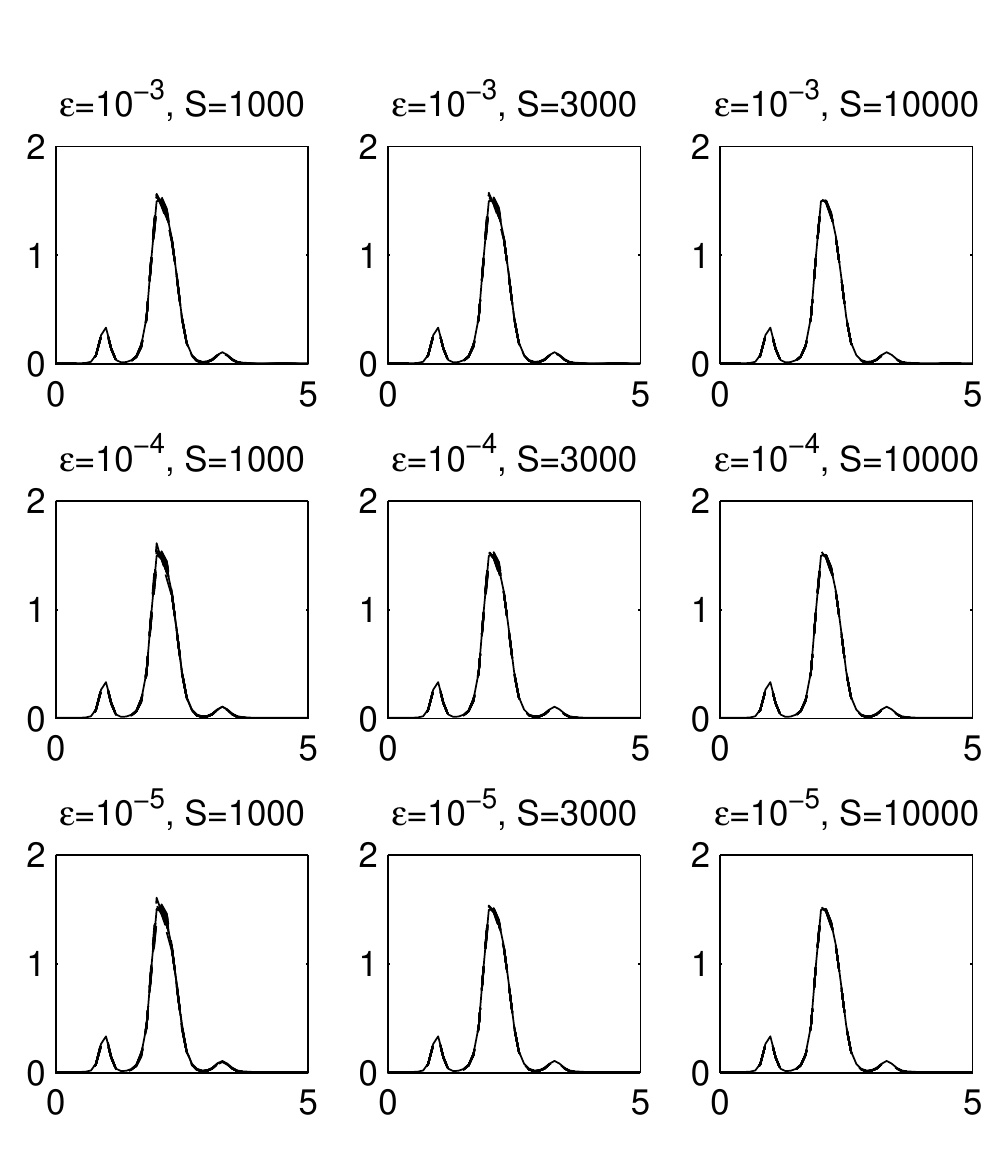}
\end{center}
\caption{\small The posterior mean density from 20 different runs of the adaptive truncation algorithm with RSB truncation of a Dirichlet process with $M=1$ using $m=3$,  and different values of $S$ and $\epsilon$ for the galaxy data. }\label{f:galaxy_plots}
\end{figure}
Initially, the problem of density estimation with $M=1$ was considered. The results of 20 different runs of the adaptive truncation algorithm using the RSB truncation with $m=3$, different numbers of particles $S$ and different values of $\epsilon$ 
 are shown in Figure~\ref{f:galaxy_plots}. The approximations of the posterior mean density improved as $S$ increases with smaller variability in the estimates but the effect of $\epsilon$ seemed negligible.
These patterns were confirmed by calculating the mean integrated  squared error (MISE) which measures the discrepancy between
 the approximations from the adaptive truncation algorithm and the infinite-dimensional posterior for different combinations of $S$ and $\epsilon$. The MISE is defined for fixed $S$ and $\epsilon$ as 
\[
\mbox{MISE} = \frac{1}{B}\sum_{i=1}^B\int \left(f^{(i)}(x) - f^{GS}(x)\right)^2 \,dx
\]
where  $B$ is the number of runs of the algorithm, $f^{(i)}(x)$ is the posterior mean density calculated using the output from the $i$-th run of the algorithm and 
$f^{GS}(x)$ is  a ``gold-standard'' estimate from the infinite dimensional posterior. 
The MISE will be small if 
the density from the infinite-dimensional posterior is well approximated across all runs of the algorithm. 
\begin{table}[h!]
\begin{center}
\begin{tabular}{cccc}\hline
                                 & $S=1000$ & $S=3000$ & $S=10000$  \\\hline
$\epsilon=10^{-3}$ & $3.32 \times 10^{-4}$  & $2.47 \times 10^{-4}$ & $1.10 \times 10^{-4}$ \\
$\epsilon=10^{-4}$ & $3.83 \times 10^{-4}$  & $2.48 \times 10^{-4}$  & $1.37 \times 10^{-4}$ \\
$\epsilon=10^{-5}$ & $4.17 \times 10^{-4}$  & $1.62 \times 10^{-4}$ & $1.22\times 10^{-4}$ \\
$\epsilon=10^{-6}$ & $3.97 \times 10^{-4}$  & $2.20 \times 10^{-4}$  & $1.10 \times 10^{-4}$ \\\hline
\end{tabular}
\end{center}
\caption{\small The mean integrated squared error (MISE) over 20 runs of the adaptive truncation algorithm with RSB truncation of a Dirichlet process with $M=1$ using $m=3$, and  different values of $S$ and $\epsilon$ for the galaxy data. }\label{t:galaxy_AISE}
\end{table}
The
 slice sampler of \cite{KaGrWa08}
was run with a burn-in of 50\ 000 iterations with a subsequent run length of 5\ 000\ 000 iterations  as the gold-standard estimate. The results for the MISE with different values of $S$ and $\epsilon$ are shown in Table~\ref{t:galaxy_AISE}. There were only small differences between the results with different values of  $\epsilon$ for fixed values of $S$ but 
the approximations improved as $S$ increased  for fixed $\epsilon$  (as we would expect).

A more challenging problem is inference about the hyperparameter $M$ of the Dirichlet process. Many truncation results previously developed in the literature assume a fixed value of $M$ and so do not easily generalize to this more complicated inference problem. 
\begin{table}[h!]
\begin{center}
\begin{tabular}{ccccccc}\hline
& \multicolumn{3}{c}{RSB} & \multicolumn{3}{c}{FK}\\\hline
                                 &  $\E[M\vert y]$ & $R$ & Time &  $\E[M\vert y]$ & $R$ & Time\\\hline
                      MCMC & 0.850 &  & & 0.850 &\\
$\epsilon=10^{-3}$ & 0.846 (0.024) & 21.6 (2.9) & 11.6 &  0.874 (0.014) & 15.8 (0.5) & 7.4\\
$\epsilon=10^{-4}$ & 0.840 (0.020) & 27.1 (3.0) & 11.9  & 0.877 (0.020) & 17.6 (0.8) & 8\\
$\epsilon=10^{-5}$ & 0.848 (0.024) & 35.4 (3.5) & 11.8 & 0.878(0.022) &  19.8 (0.4)& 8.8\\
$\epsilon=10^{-6}$ & 0.842 (0.025) & 43.9 (4.4) & 11.6 & 0.863 (0.032) & 22.0 (0.7) & 9.7\\\hline
\end{tabular}
\end{center}
\caption{\small Summaries of the estimated posterior mean of $M$ and the stopping time $R$ of the algorithm and computational time in minutes  over 20 runs of the adaptive truncation algorithm with RSB truncation and FK truncation of a Dirichlet process. 
The parameter $M$ is assumed unknown and the algorithmic parameters were  $m=3$, $S=10\ 000$ and different values of $\epsilon$ for the galaxy data. The reported values are averaged over the 20 runs with sample standard deviations shown in brackets.}\label{t:galaxy_DP_M}
\end{table}
The adaptive truncation algorithm with RSB truncation was run using 10\ 000 particles, $m=3$, and different values of $\epsilon$. The hyperparameter $M$ was given an exponential prior with mean 1.
Some results for posterior inference about $M$, the stopping time $R$ and the computational time are given in Table~\ref{t:galaxy_DP_M}.
The ``MCMC'' results were calculated using the slice sampling algorithm described in \cite{KaGrWa08} 
which was run for the same number of iterations as the GS approximation used to calculate MISE.
The slice sampler generates samples from the infinite-dimensional prior and so allowed quantification of the truncation error for  different values of $\epsilon$. The adaptive truncation algorithm gave estimates of $\E[M\vert y]$ which are very similar to those from the infinite dimensional posterior for all values of $\epsilon$. Table~\ref{t:galaxy_DP_M} also shows the results using the FK truncation with 10\ 000 particles. These typically had a larger error (although, the error was still not particularly large). As we would expected the stopping time increases on average with $T$ for both the RSB and FK truncations. This leads to clearly increasing running times for the FK truncation but the effect is not clear with the RSB truncation where running times are very similar. This is due to the structure of the algorithm where computational effort is divided between running the MCMC sampler for $\pi_1(\cdot\vert y)$ and sequentially proposing from the transition density $\pi_{k+1}(\phi_{k+1}\vert \phi_{1:k},\kappa)$. Difference in the stopping time need not  have a large effect on overall computational time if the transition density can be sampled quickly relative to the MCMC sampler. The results are consistent with this observation. The transitions in the RSB truncation involves sampling a single beta random variables whereas the transition in the FK
 truncation involves a numerical inversion to find the next value in the Ferguson-Klass representation. The level of truncation error suggests that the RSB truncation should be preferred to the FK truncation in this problem.

The Dirichlet process has weights which decay exponentially in the SB representation. Other specifications of the nonparametric prior lead to a slower decay of the weights. One such prior is the Poisson-Dirichlet process 
\citep{PiYo97}. The rate of decay of the weights decreases as $a$ increases and large $a$ is associated with very slow decay of weights. This is an important test case for truncation methods since the slow decay of the weights can lead to large truncation errors unless many atoms are included in the approximation. The adaptive truncation algorithm was tested 
on  the infinite mixture model described at the start of section where the Poisson-Dirichlet process was used as the nonparametric prior. The parameters of the process were considered unknown. The parameter $a$ was given a uniform prior on $(0,1)$ and  $M$ was given an exponential prior with mean 1. 
\begin{table}[h!]
\begin{center}
\begin{tabular}{ccccc}\hline
                                 & $\E[a\vert y]$ & $\E[M\vert y]$ & $R$ & Time\\\hline
                      MCMC & 0.193 & 0.591 & &\\
$\epsilon=10^{-3}$ & 0.219 (0.004) & 0.569 (0.011) & 62.2 (7.3) & 66\\
$\epsilon=10^{-4}$ & 0.203 (0.003) & 0.574 (0.012) & 154.8 (16.3) & 85 \\
$\epsilon=10^{-5}$ & 0.198 (0.004) & 0.577 (0.017) & 475.6 (73.9) & 106\\
$\epsilon=10^{-6}$ & 0.196 (0.002) & 0.573 (0.017) & 1603.3 (319.7) & 124\\\hline
\end{tabular}
\end{center}
\caption{\small 
 Summaries of the estimated posterior mean of $a$ and $M$, the stopping time $R$ of the algorithm and computational time in minutes 
 over 20 runs of the adaptive truncation algorithm wtih RSB truncation of a Poisson-Dirichlet process. The parameters $a$ and $M$ were assumed unknown and the algorithmic parameters were $m=3$, $S=10\ 000$ and different values of $\epsilon$ for the galaxy data. The reported values are averaged over the 20 runs with sample standard deviations shown in brackets.}\label{t:galaxy_PY}
\end{table}
Some results from the adaptive truncation algorithm with RSB truncation run with $m=3$ and 10\ 000 particles
 are given  in Table~\ref{t:galaxy_PY}. The ``MCMC'' results were calculated using the method described for the 
Dirichlet process mixture model. In this case, the value of $\epsilon$ had some impact on the quality of approximation.
 The truncation error in estimating the posterior mean of $a$ and $M$ decreased as $\epsilon$ decreased as we would expect. The estimates with $\epsilon=10^{-3}$ are close to the MCMC results and become very close for $\epsilon=10^{-5}$ and $\epsilon=10^{-6}$ illustrating that the adaptive truncation algorithm can work well in this challenging example. The price to be paid for the increased accuracy is typically  larger stopping times which increases from 62.2  for $\epsilon=10^{-3}$ to to 1603.3 for $\epsilon=10^{-6}$ and much longer computational times.



\subsection{A semiparametric linear mixed  model}

Linear mixed models are a popular way to model the heterogeneity of subjects with repeated measurements. 
It is assumed that responses  $y_{i1},\dots,y_{iT}$ are observed for the $i$-th subject with $(1\times p)$-dimensional vectors of regressors $X_{i1},\dots,X_{iT}$ 
and $(1\times q)$-dimensional vectors of regressors  $Z_{i1},\dots,Z_{iT}$. The vectors $X_{it}$ and $Z_{it}$ may have some elements  in common. The usual linear mixed effects model assumes assumes that 
\begin{equation}
y_{it}= X_{it}\beta +  Z_{it}\gamma_{i} + \epsilon_{it},\qquad 
i=1,\dots,n,\quad t=1,\dots,T
\label{lmm1}
\end{equation}
where $\beta$ is a $(p\times 1)$-dimensional vector of  fixed effects and $\gamma_i$ is  $(q\times 1)$-dimensional vector of   random effects for the $i$-th subject. The model is usually made identifiable by assuming that $\E[\gamma_{i}]=0$ and $\E[\epsilon_{it}]=0$, which implies that  $\E[y_{it}\vert X_{it},Z_{it}]=X_{it}\beta$ and allows the regression effects $\beta$ to be interpreted in the same way as in a linear regression model. Often parametric distributions are chosen for the errors  and the random effects  with  $\epsilon_{it}\sim\N(0,\sigma^2)$ and $\gamma_i\sim\N(0,\Sigma_{\gamma})$ being standard choices. However, in general, little is often known {\it a priori} about the distribution of the errors or the random effects and many authors have argued for a nonparametric approach. 

Bayesian nonparametric inference in linear mixed models was initially considered by \cite{BuMa96} and \cite{KlIb98} and subsequently developed by \cite{IshTak02}. These models assume that $\gamma_i$ is given a Dirichlet process mixture prior but use a parametric distribution for the errors. The mean of a Dirichlet process is a random variable and so the condition that $\E[\gamma_i]=0$ is not imposed on the model. Typically, an alternative model
is used where 
\begin{equation}
y_{it}=X^{\star}_{it}\beta^{\star} + Z_{it}\gamma_i^{\star} + \epsilon_{it},\qquad 
i=1,\dots,n,\quad t=1,\dots,T
\label{lmm2}
\end{equation}
and it is assumed that all elements of $Z_{it}$ appear in $X_{it}$ and that $X_{it}^{\star}$ is defined to be a design matrix containing the elements of $X_{it}$ not shared with $Z_{it}$. This removes the need for the identifiability constraint on the random effect since (\ref{lmm1}) implies that $\E[\gamma_i^{\star}]$ can be non-zero.  \cite{LiMuLi10} discuss using post-processing of MCMC samples from (\ref{lmm2}) to make inference about the parameters in (\ref{lmm1}).

An alternative approach to inference in linear mixed models directly imposes location constraints on the nonparametric prior. 
\cite{KoGe01} and \cite{HaJo02} constructed error distribution with median zero using mixtures of uniforms and P\'olya tree priors respectively. \cite{Tok06} constructed nonparametric priors whose realizations are zero mean distributions using the symmetrized Dirichlet process. I considered imposing constraints on  the nonparametric priors for $\epsilon_{it}$ and $\gamma_i$ so that $\E[\epsilon_{it}]=0$ and  $\E[\gamma_i]=0$ using the method of \cite{YanDunBai10}. They assumed that $\epsilon_{it}=\tilde{\epsilon}_{it}-\E[\tilde{\epsilon}_{it}]$  and $\gamma_i=\tilde{\gamma}_i-\E[\tilde{\gamma}_i]$ 
where $\tilde\epsilon_{it}$ and $\gamma_i$ are given nonparametric priors without a mean constraint. 

I assumed that $q=1$ and used versions of the CCV model \citep{Gri10} as the nonparametric priors,
\[
p\left(\tilde\epsilon_{it}\right)=\sum_{j=1}^{\infty} p_j^{\epsilon}\,\N\left(\tilde\epsilon_{it}\left\vert \mu^{\epsilon}_j,a_{\epsilon}\sigma_{\epsilon}^2\right.\right)\mbox{ and }
p\left(\tilde\gamma_{i}\right)=\sum_{j=1}^{\infty} p_j^{\gamma}\,\N\left(\tilde\gamma_{i}\left\vert \mu^{\gamma}_j,a_{\gamma}\sigma_{\gamma}^2\right.\right)
\]
where the Dirichlet process prior for $\left(p^{\epsilon}_1,\mu_1^{\epsilon}\right), \left(p^{\epsilon}_2,\mu_2^{\epsilon}\right), \dots$ has mass parameter $M_{\epsilon}$ and 
centring measure  $\N\left(0,(1-a_{\epsilon})\sigma_{\epsilon}^2\right)$.
Similarly,  the Dirichlet process prior for $\left(p^{\gamma}_l,\mu_l^{\gamma}\right)$ has mass parameter $M_{\gamma}$ and 
centring measure $\N\left(\vert 0,(1-a_{\gamma})\sigma_{\gamma}^2\right)$.
This allows $\sigma_{\epsilon}$ and $\sigma_{\gamma}$ to be interpreted as scale of $\epsilon_{it}$ and $\gamma_i$ respectively. The parameters $a_{\epsilon}$  and $a_{\gamma}$ are smoothness parameters with smaller values of indicating a rough density with potentially more modes. \cite{YanDunBai10} used a truncated version of the Dirichlet process prior to fit these types of models but do not develop any specific theory for choosing the number of atoms. I will consider using the adaptive truncation algorithm with RSB truncation to avoid choosing this value before running any algorithms. Details of the algorithm are given in Appendix A.1.

\begin{figure}[h!]
\begin{center}
\includegraphics[trim=0mm 0mm 90mm 230mm, clip, scale=1]{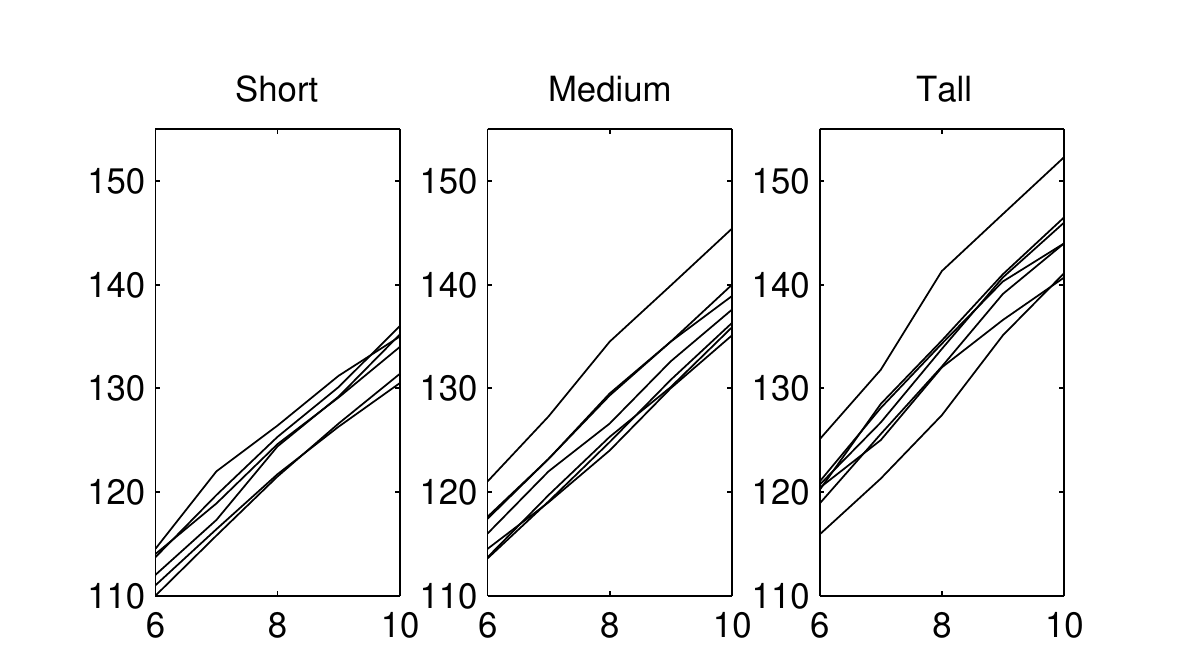}
\end{center}
\caption{\small The observed growth curves in the three groups in the schoolgirl data.}\label{f:height_data}
\end{figure}
The method will be illustrated on the ``schoolgirl'' data set taken from the ``DPpackage'' in R. The data are the heights of 20 children measured at ages from 6 to 10 inclusive and the height of their mothers which were divided into three groups (short, medium or tall). The data are shown in Figure~\ref{f:height_data}.
The groups were included using dummy variables and age was included as a regressor. The intercept for each schoolgirl was assumed to be a random effect. This leads to $n=20$, $T=5$, $p=4$ and $q=1$ with $Z_{it1}=1$ for all $i$ and $t$.
The nonparametric model was fitted to the data with
 the following hyperpriors:
$a_{\epsilon}\sim\Be(1,19)$, $\sigma^2_{\epsilon}\sim\mbox{FT}(1, 0.01)$, $M_{\epsilon}\sim\Ga(1,1)$, $a_{\gamma}\sim\Be(1,19)$, $\sigma^2_{\gamma}\sim\mbox{FT}(1, 1)$ and $M_{\gamma}\sim\Ga(1,1)$,
$\beta\sim\N(0,10^6 I_4)$. The notation $\mbox{FT}(\nu,A)$ denotes the folded $t$-distribution which has 
the density
\[
p(x)=\left(1+\frac{x}{A}\right)^{-(\nu+1)/2},\qquad x>0.
\]
This is  a heavy-tailed distribution which was proposed by \cite{gel06} for variance parameters in hierarchical models. 
The adaptive truncation algorithm was run with 10\ 000 particles, $\epsilon=10^{-5}$ and $m=3$. The initial MCMC run to simulate values from $\pi_1(\phi_1,\lambda,\kappa\vert y)$ used a burn-in period of 5\ 000 iteration with a thinning of every fifth value.
\begin{figure}[h!]
\begin{center}
\includegraphics[trim=0mm 0mm 90mm 230mm, clip, scale=1]{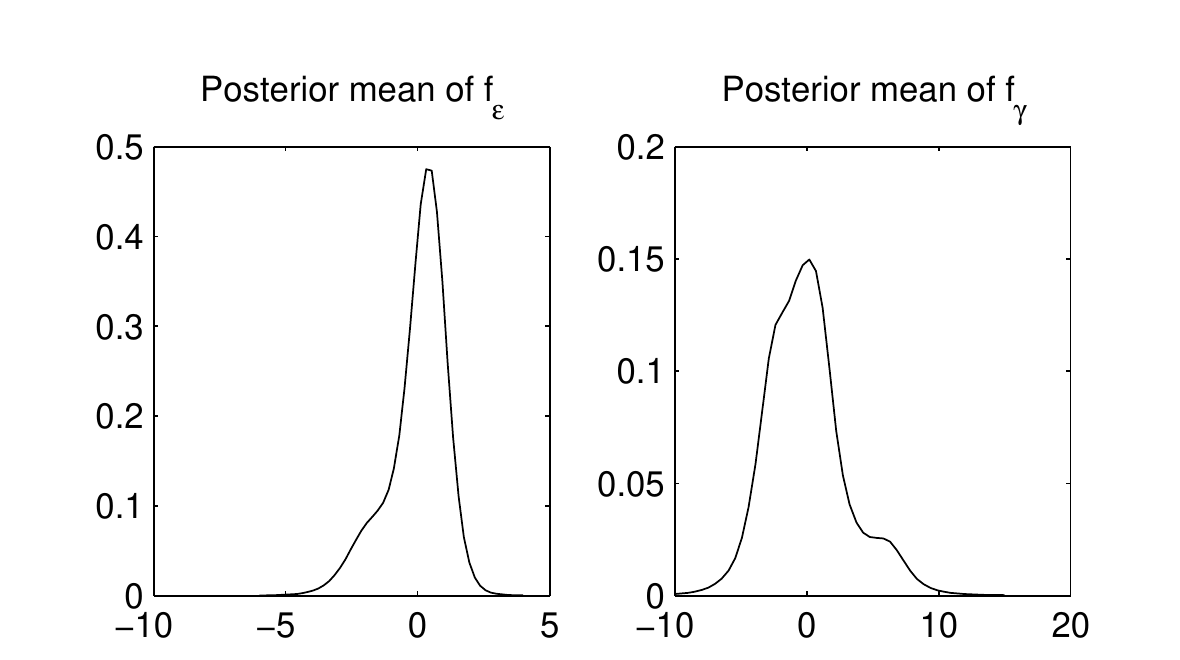}
\end{center}
\caption{\small The posterior mean of the density of the observation error, $f_{\epsilon}$, and the random effect, 
$f_{\gamma}$ with the schoolgirl data.}\label{f:height_density}
\end{figure}
The densities of the observational error, $f_{\epsilon}$, and the random effect, $f_{\gamma}$, were summarized by their posterior means which are shown 
in Figure~\ref{f:height_density}. In both cases, the densities clearly deviated from normality. The posterior mean of $f_{\epsilon}$ 
had a clear negative skewness and the posterior mean of $f_{\gamma}$ had a positive skewness. 
\begin{figure}[h!]
\begin{center}
\includegraphics[trim=0mm 0mm 90mm 230mm, clip, scale=1]{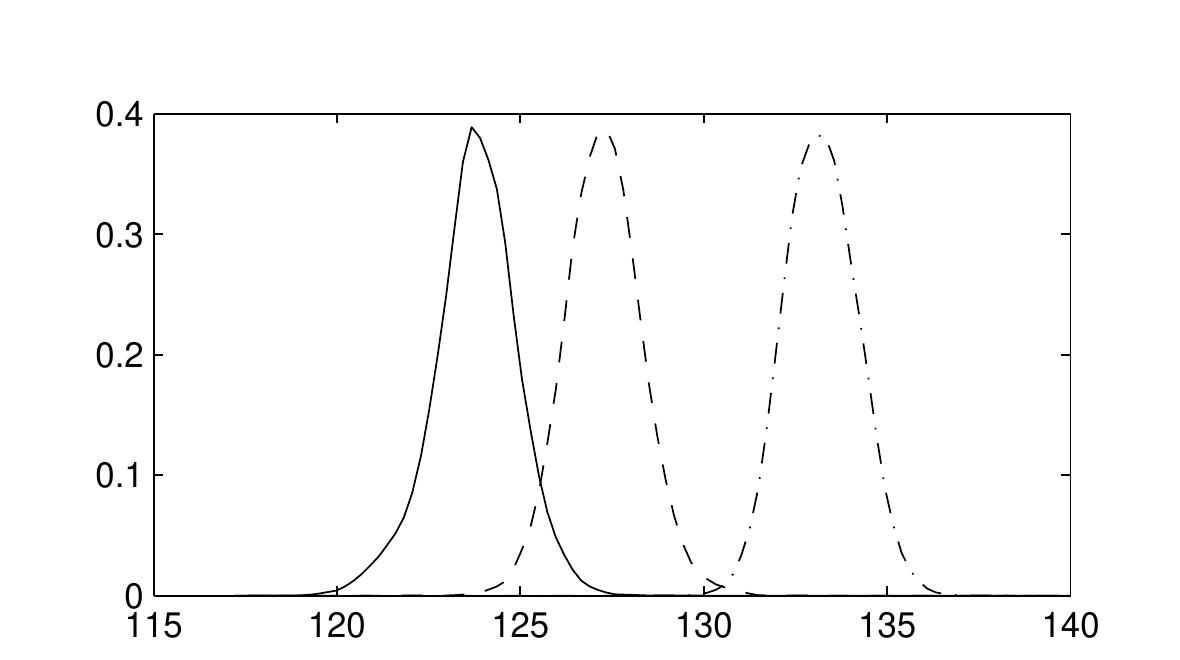}
\end{center}
\caption{\small The posterior densities of the group means for: the short group (solid line), medium group (dashed line) and the tall group (dot-dashed line) with the schoolgirl data.}\label{f:height_beta}
\end{figure}
The posterior densities of the group means are illustrated in Figure~\ref{f:height_beta}. These show some evidence of differences between the group means with the largest differences between the tall group and the other two groups with a much less marked difference between the small and medium group means.

\subsection{A nonparametric  time series model}

\cite{AntWal12} described a method for Bayesian nonparametric inference in stationary time series models. Suppose that $z_1,\dots,z_T$ are a stationary time series, their model assumes that  the transition probability is
\[
p(z_t\vert z_{t-1})=\frac{\sum_{j=1}^{\infty} p_j \,\N\left(\left.\left(\begin{array}{c}z_{t-1}\\ z_t\end{array}\right)\right\vert \left(\begin{array}{c}\mu_j\\\mu_j\end{array}\right),\sigma_i^2\left(\begin{array}{cc} 1 & \rho_i\\
\rho_i & 1
\end{array}
\right)
\right)}
{\sum_{j=1}^{\infty} p_j \,\N(z_{t-1}\vert \mu_j,\sigma_j^2)}
\]
and that the distribution of the initial value is
\[
p(z_1)=\sum_{j=1}^{\infty} p_j \,\N(z_{t-1}\vert \mu_j,\sigma_j^2).
\]
The stationary distribution is 
$\sum_{j=1}^{\infty} p_j \,\N(z_t\vert \mu_j,\sigma_j^2)$ and the nonparametric specification of the transition density allows dependence to be flexibly modelled. Bayesian nonparametric inference in this model involves placing a prior on
$G_z=\sum_{j=1}^{\infty}p_j\delta_{\mu_j,\sigma_j,\rho_j}$
 and \cite{AntWal12}  show that the prior has large support if $G_z$ is given a Dirichlet process prior.

In practice, many observed data series are non-stationary. A simple model for a non-stationary time series $y_1,\dots,y_T$ has the form 
\[
y_t = \alpha_t + \epsilon_t,\qquad t=1,\dots,T
\]
where $\alpha_t$ is a random walk component and $\epsilon_t$ is a stationary process component. A flexible specification of this model would assume that $\alpha_t$ follows a random walk whose increments are drawn from a unknown distribution
and $\epsilon_t$ follows the nonparametric model of \cite{AntWal12}. The process for $\alpha_t$ is
\[
 \alpha_t=\alpha_{t-1}+\nu_t, \qquad p(\nu_t)=\sum_{j=1}^{\infty} p^{\alpha}_j\,\N\left(\nu_t\left\vert \mu_j^{\alpha},a_{\alpha}\sigma^2_{\alpha}\right.\right)
\] 
where $(p_j^{\alpha},\mu_j^{\alpha})$ are given a Dirichlet process prior with mass parameter $M_{\alpha}$
and centring measure
 $\N\left(0, (1-a_{\alpha})\sigma^2_{\alpha}\right)$.
The stationary process $\epsilon_t$ is given a variation on the nonparametric  prior described by \cite{AntWal12} which assumes that $\epsilon_t=\tilde\epsilon_t-\E[\tilde\epsilon_t]$,
$\sigma^2_i=a_{\epsilon}\sigma^2_{\epsilon}$ and $G_{\epsilon}\sim\DP(M_{\epsilon}H_{\epsilon})$ where $H_{\epsilon}$ is a zero-mean normal distribution with variance 
$(1-\alpha_{\epsilon})\sigma^2_{\epsilon}$. This ensures that the stationary distribution of $\epsilon_t$ has zero expectation. Details of the algorithm are given in Appendix A.2.

\begin{figure}[h!]
\begin{center}
\includegraphics[trim=10mm 0mm 100mm 220mm, clip, scale=1]{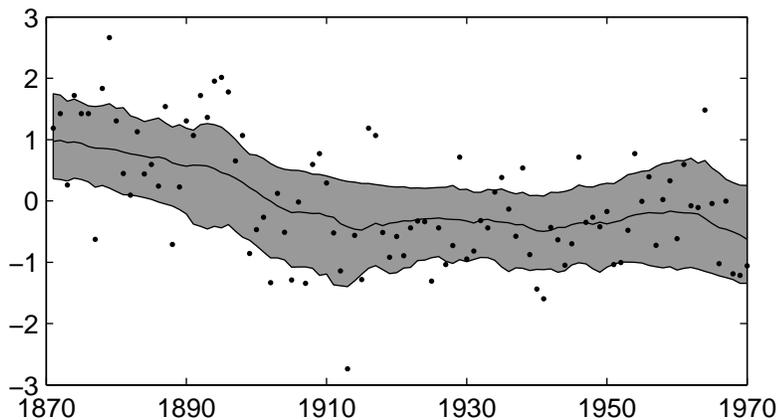}
\end{center}
\caption{\small The data and the posterior median of $\alpha_t$ (solid line) with point-wise 95\% credible interval for the Nile flow data.}
\label{f:Nile_data}
\end{figure}
As an illustration, the model was applied to measurements of the annual flow of the Nile at Ashwan from 1871 to 1970, which is available from the ``datasets''  package in R. The data were standardized by subtracting the mean and dividing by the standard deviation and  are plotted in Figure~\ref{f:Nile_data}. The graph  shows clear evidence of non-stationary with a higher average level of flow in the initial years of the sample. The nonparametric model was fitted to the data with
 the following hyperpriors:
$a_{\alpha}\sim\Be(1,19)$, $\sigma^2_{\alpha}\sim\mbox{FT}(1, 0.01)$, $M_{\alpha}\sim\Ga(1,1)$, $a_{\epsilon}\sim\Be(1,19)$, $\sigma^2_{\epsilon}\sim\mbox{FT}(1, 1)$, $M_{\epsilon}\sim\Ga(1,1)$ and 
$\alpha_1\sim\N(0,1)$. The values of $\sigma^2_{\alpha}$ and $\sigma^2_{\epsilon}$ were centred over values which allow all the variation in the data to be explained by one of the components, which represents a conservative choice.
The adaptive truncation algorithm with RSB truncation was run using 10\ 000 particles, $\epsilon=10^{-5}$ and $m=3$. The initial MCMC run to simulate values from $\pi_1(\phi_1,\lambda,\kappa\vert y)$ 
used a burn-in period of 10\ 000 with a thinning of every fifth value.
The posterior mean 
of the trend $\alpha_t$ is plotted in Figure~\ref{f:Nile_data}. This clearly shows that the average flow of the Nile fell over the initial period of the data.
\begin{figure}[h!]
\begin{center}
\begin{tabular}{cc}
\includegraphics[trim=0mm 0mm 90mm 230mm, clip, scale=1]{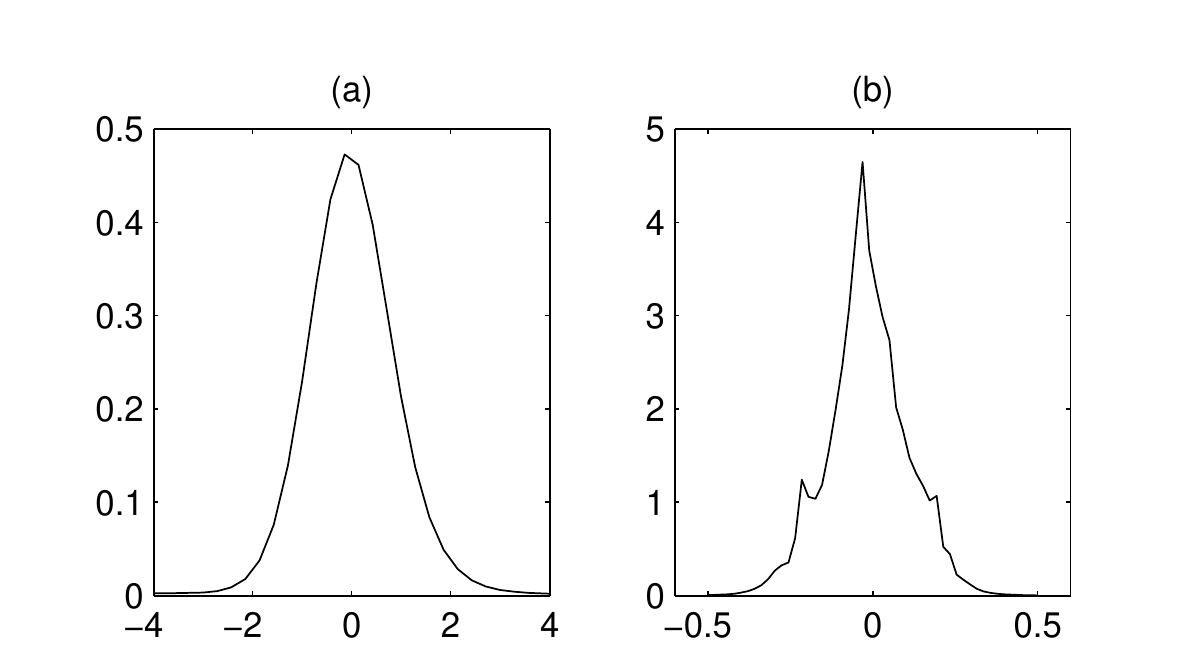}
\end{tabular}
\end{center}
\caption{\small The posterior mean density of: (a) $\nu_t$ and (b) $\epsilon_t$ for the Nile flow data.}\label{f:Nile_inc}
\end{figure}
The posterior mean stationary density of $\epsilon_t$ and the posterior mean density of $\nu_t$  are shown in Figure~\ref{f:Nile_inc}. The posterior mean stationary density of $\epsilon_t$ had a slight positive skew and 
the posterior mean density of $\nu_t$ is heavy tailed with a pronounced spike at the mode.

\begin{figure}[h!]
\begin{center}
\includegraphics[trim=0mm 0mm 90mm 230mm, clip, scale=1]{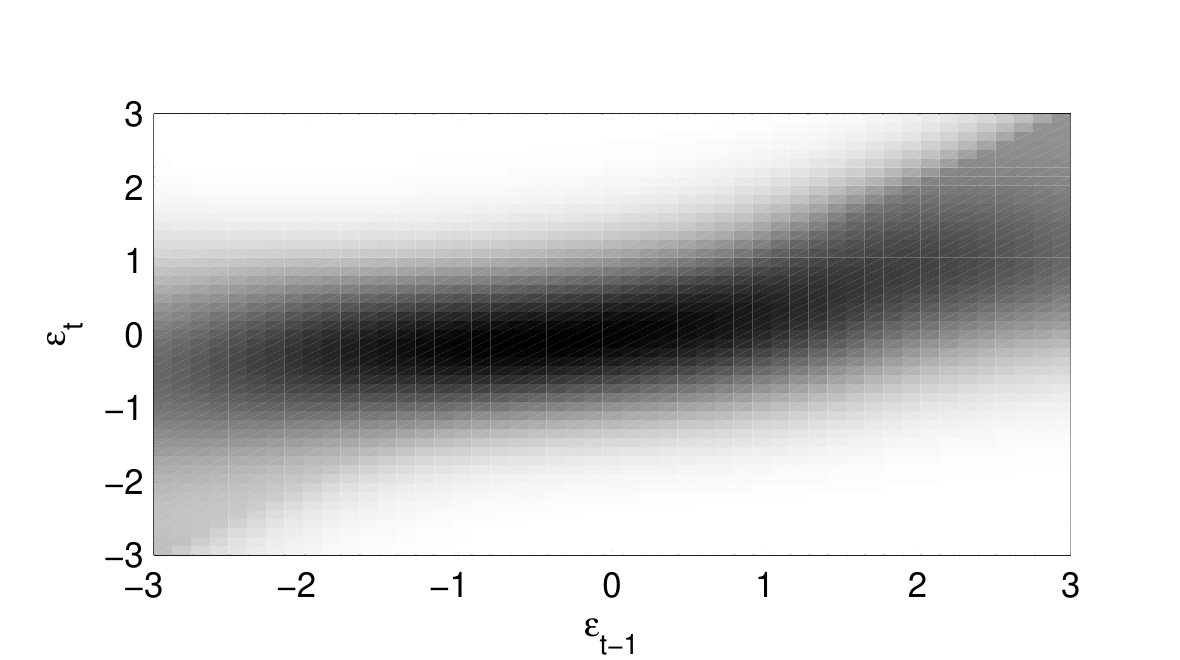}
\end{center}
\caption{\small A heat map of the posterior mean transition density of the stationary component $\epsilon_t$ for the Nile flow data. Darker colours represent higher density values.}\label{f:Nile_trans}
\end{figure}
The transition density of the stationary component $\epsilon_t$ is shown in Figure~\ref{f:Nile_trans}. There was clear evidence of a departure from the assumptions of an AR(1) process (which would be represented by diagonal bands with the same colour). The transition density was  negatively skewed for values of $\epsilon_{t-1}$ less than 0 whereas the density was roughly symmetric for $\epsilon_t$ greater than 0. The conditional mean of $\epsilon_{t}$ increased more quickly with $\epsilon_{t-1}$ for positive $\epsilon_{t-1}$ compared to negative $\epsilon_{t-1}$.

\section{Discussion}

This paper descibes a method for adaptively choosing the truncation point for posterior inference in nonparametric models.
Application to the infinite mixture models showed that these methods can be effective for both density estimation and inference about hyperparameters of the nonparametric prior. The adaptive truncation method can be easily applied to non-standard mixture models, such as those fitted in Sections 5.2 and 5.3, which cannot be fitted 
with the infinite-dimensional prior
using MCMC methods 

The methods developed in this article have relatively simple proposals in the SMC steps and only update global parameters if Step 4 occurs. This works well in the examples considered here but the current method has potential for further development which may be particularly important for problems where the nonparametric prior is defined on a high-dimensional space. These include variation on the general proposal mechanisms described by \cite{DMDoJa06} and the generic SMC$^2$ method \citep{ChJaPa13} which allow generalization of type of SMC methods (with MCMC steps) used in the adaptive truncation algorithm.

This paper has concentrated on inference in mixture models with nonparametric priors which is the most popular class of models in Bayesian nonparametric modelling. The adaptive truncation method is generic and can be applied to a much wider class of models. One increasingly important class of models are latent variable models with an infinite number of latent variables or processes. Examples include infinite factor models \citep{BhDu11}, infinite aggregation models \citep{KaGr11}, and linear models with L\'evy process priors \citep{PoSc12}. Future work will consider the application of the adaptive truncation algorithm to these models.

\bibliographystyle{chicago}
\bibliography{ref_work}

\appendix
\section{Samplers}

\subsection{A semiparametric linear mixed model}

The mixture distributions in the model are approximated using the RSB truncation leading to the $k$-th truncated distributions
\[
p\left(\tilde\epsilon_{it}\right)=\sum_{j=1}^{N_k} p_j^{\epsilon}\,\N\left(\tilde\epsilon_{it}\left\vert \mu^{\epsilon}_j,a_{\epsilon}\sigma_{\epsilon}^2\right.\right)\mbox{ and }
p\left(\tilde\gamma_{i}\right)=\sum_{j=1}^{N_k} p_j^{\gamma}\,\N\left(\tilde\gamma_{i}\left\vert \mu^{\gamma}_j,a_{\gamma}\sigma_{\gamma}^2\right.\right)
\]
where $p_j^{\epsilon}=\frac{V^{\epsilon}_l\prod_{l<j}\left(1-V^{\epsilon}_l\right)}{1-\prod_{l=1}^{N_k} (1-V^{\epsilon}_l)}$ and $p_j^{\gamma}=\frac{V^{\gamma}_j\prod_{l<j}\left(1-V^{\gamma}_l\right)}{1-\prod_{l=1}^{N_k} (1-V^{\gamma}_l)}$
 with $V^{\epsilon}_j\sim\Be(1,M_{\epsilon})$ and $V^{\gamma}_j\sim\Be(1,M_{\gamma})$
 for $j=1,\dots,N_k$.
 The parameters in the initial truncation are $\phi_1=\left(V^{\epsilon}_{1:N_1}, \mu^{\epsilon}_{1:N_1},V^{\gamma}_{1:N_1}, \mu^{\gamma}_{1:N_1}\right)$
and the extra parameters added to form the $k$-th truncation are 
 $\phi_k=\left(V^{\epsilon}_{N_k}, \mu^{\epsilon}_{N_k},V^{\gamma}_{N_k}, \mu^{\gamma}_{N_k}\right)$. The algorithm introduces allocation variables $s^{\epsilon}_{it}$ and $s^{\gamma}_i$ for $\epsilon_{it}$ and $\gamma_i$ respectively
with $i=1,\dots,n$ and $j=1,\dots,T$. These are defined by
\[
p(s_{it}^{\epsilon}=j)=p_j^{\epsilon}\mbox{ and }p(s_{i}^{\gamma}=j)=p_j^{\gamma},\qquad j=1,\dots,N_k.
\]
It is useful to define
\begin{align*}
&f_k(a_{\epsilon},\sigma^2_{\epsilon},a_{\gamma},\sigma^2_{\gamma},\beta,\mu^{\epsilon},V^{\epsilon},\mu^{\gamma},V^{\gamma})\\
&=
\exp\left\{-\frac{1}{2}\left[\frac{\sum_{i=1}^n\sum_{t=1}^Td_{it}^{\epsilon\,2}}{a_{\epsilon}\sigma^2_{\epsilon}}
+\frac{\sum_{i=1}^n d_i^{\gamma\,2}}{a_{\gamma}\sigma^2_{\gamma}}
- \frac{\sum_{i=1}^n c_i^2}{d}
\right]
\right\}
(a_{\gamma}\sigma^2_{\gamma})^{-n/2}
(a_{\epsilon}\sigma^2_{\epsilon})^{-nT/2}\\
&\quad \times d^{-n/2}
\end{align*}
where $d_{it}^{\epsilon}=y_{it} - X_{it}\beta - \mu^{\epsilon}_{s^{\epsilon}_{it}} + \bar{\mu^{\epsilon}}$, 
$d_{i}^{\gamma}=\mu^{\gamma}_{s^{\gamma}_i} - \bar{\mu^{\gamma}}$,
$c_i=
\frac{\sum_{t=1}^T d^{\epsilon}_{it}}{a_{\epsilon}\sigma^2_{\epsilon}}+
\frac{d^{\gamma}_i}{a_{\gamma}\sigma^2_{\gamma}}$,
and $d=\frac{T}{a_{\epsilon}\sigma^2_{\epsilon}}+\frac{1}{a_{\gamma}\sigma^2_{\gamma}}$
and the values are calculated for the $k$-th truncation. 
Since $s^{\gamma}$ and $s^{\epsilon}$ are included in the sampling, values are proposed in Step 1 of the adaptive truncation algorithm.
At the $k$-th iteration, the transition for the allocation $s^{\gamma}_{i}$ is
\[
s^{\gamma}_{i} = \left\{
\begin{array}{ll}
N_k & \mbox{with probability }1-r^{\gamma},\\
s^{\gamma}_{i}  & \mbox{otherwise}
\end{array}
\right.
\]
and the transition for the allocation $s^{\epsilon}_{i,t}$ is 
\[
s^{\epsilon}_{i,t} = \left\{
\begin{array}{ll}
N_k & \mbox{with probability }1-r^{\epsilon}\\
s^{\epsilon}_{i,t}  & \mbox{otherwise}
\end{array}
\right.
\]
where $r_{\gamma}=\frac{1-\prod_{j=1}^{N_k} \left(1-V_j^{\gamma}\right)}
{1-\prod_{j=1}^{N_k} \left(1-V_j^{\gamma}\right)+p^{\gamma}_{N_k}}$
and $r_{\epsilon}=\frac{1-\prod_{j=1}^{N_k} \left(1-V_j^{\epsilon}\right)}
{1-\prod_{j=1}^{N_k} \left(1-V_j^{\epsilon}\right)+p^{\epsilon}_{N_k}}$. 
In this case, the weights in Step 2 are updated using
\[
\alpha_k^{(j)}=
\frac{f_{k+1}\left(a^{(j)}_{\epsilon},\sigma^{2\,(j)}_{\epsilon},a^{(j)}_{\gamma},\sigma^{2\,(j)}_{\gamma},\beta^{(j)},\mu^{\epsilon\,(j)},V^{\epsilon\,(j)},\mu^{\gamma\,(j)},V^{\gamma\,(j)}\right)}
{f_k\left(a^{(j)}_{\epsilon},\sigma^{2\,(j)}_{\epsilon},a^{(j)}_{\gamma},\sigma^{2\,(j)}_{\gamma},\beta^{(j)},\mu^{\epsilon\,(j)},V^{\epsilon\,(j)},\mu^{\gamma\,(j)},V^{\gamma\,(j)}\right)}.
\]

The MCMC sampler updates many parameters using a variation of the adaptive Metropolis-Hastings random walk algorithm of
\cite{atros05} which  allows the proposal density to be updated at each iteration of the sampler.
It works as follows for a generic parameter value $\tau$. Let $q^{(i)}(\tau,\tau')=\N(\tau'\vert \tau,\sigma^{2\,(i)}_{\tau})$
 be the proposal
 density for $\tau$ at the $i$-th iteration which is random walk proposal with variance $\sigma^{2\,(i)}_{\tau}$
and let $\alpha_i$ be the acceptance probability of a Metropolis-Hastings move with this proposal. The proposed value $\tau'$ is accepted or rejected in the standard way for a Metropolis-Hastings step.  The variance of the proposal is 
updated in the following way
\[
\log\sigma^{2\,(i+1)}_{\tau}=
\rho\left(\log\sigma^{2\,(i)}_{\tau}+i^{-c}\left(\alpha_i - \hat{\alpha}\right)\right)
\]
where $0.5<c\leq 1$ (the value $c=0.55$ was used in the examples), $\hat\alpha$ is a target acceptance rate (the conservative choice $\hat\alpha=0.3$ was used in the examples) and 
\[
\rho(x)=\left\{
\begin{array}{cc}
- b & x<-b,\\
x &  - b\leq x\leq b,\\
b & x>b
\end{array}
\right.
\]
where $b$ is taken to be very large. The average acceptance rate for the Metropolis-Hastings update of the parameter 
will converge to $\hat{\alpha}$ over the run of the sampler.
The steps of the MCMC algorithm to sample from the posterior with the $k$-th truncation are described below.

\subsubsection*{Updating $\mu^{\epsilon}_j$}

The full conditional density of $\mu^{\epsilon}_j$ is proportional to
\[
f_k\left(a_{\epsilon},\sigma^2_{\epsilon},a_{\gamma},\sigma^2_{\gamma},\beta,\mu^{\epsilon},V^{\epsilon},\theta^{\gamma},V^{\gamma}\right)
\exp\left\{-\frac{{\mu^{\epsilon}_j}^2}{(1-a_{\epsilon})\sigma^2_{\epsilon}}\right\}
\]
and the parameter is updated using the adaptive Metropolis-Hastings random walk step.

\subsubsection*{Updating $V^{\epsilon}_j$}

The full conditional density of $V^{\epsilon}_j$ is proportional to
\[
f_k\left(a_{\epsilon},\sigma^2_{\epsilon},a_{\gamma},\sigma^2_{\gamma},\beta,\mu^{\epsilon},V^{\epsilon},\mu^{\gamma},V^{\gamma}\right)
\left(1-V^{\epsilon}_j\right)^{M_{\epsilon}-1}
\]
and the parameter is updated using the adaptive Metropolis-Hastings random walk step after taking the transformation $\log V^{\epsilon}_j - \log \left(1- V^{\epsilon}_j\right)$.

\subsubsection*{Updating $M_{\epsilon}$}

The full conditional distribution of $M_{\epsilon}$ is $\Be\left(a^{\star}, b^{\star}\right)$ where $a^{\star}=1+N_k$
and $b^{\star} = 1 - \sum_{j=1}^{N_k} \log \left(1-V^{\epsilon}_j\right)$.

\subsubsection*{Updating $a_{\epsilon}$}

The full conditional density of $a_{\epsilon}$ is proportional to
\[
f_k\left(a_{\epsilon},\sigma^2_{\epsilon},a_{\gamma},\sigma^2_{\gamma},\beta,\mu^{\epsilon},V^{\epsilon},\mu^{\gamma},V^{\gamma}\right)
\exp\left\{-\frac{\sum_{j=1}^{N_k}{\mu^{\epsilon}_j}^2}{(1-a_{\epsilon})\sigma^2_{\epsilon}}\right\}
\left(1-a_{\epsilon}\right)^{18-N_k/2}
\]
and the parameter is updated using the adaptive Metropolis-Hastings random walk step after taking the transformation $\log a_{\epsilon} - \log \left(1- a_{\epsilon}\right)$.

\subsubsection*{Updating $\sigma^2_{\epsilon}$}

The full conditional density of $\sigma^2_{\epsilon}$ is proportional to
\[
f_k\left(a_{\epsilon},\sigma^2_{\epsilon},a_{\gamma},\sigma^2_{\gamma},\beta,\mu^{\epsilon},V^{\epsilon},\mu^{\gamma},V^{\gamma}\right)
\exp\left\{-\frac{\sum_{j=1}^{N_k}{\mu^{\epsilon}_j}^2}{(1-a_{\epsilon})\sigma^2_{\epsilon}}\right\}
\left(\sigma^2_{\epsilon}\right)^{-N_k/2}\left(1+\sigma^2_{\epsilon}\right)^{-1}
\]
and the parameter is updated using the adaptive Metropolis-Hastings random walk step after taking the transformation $\log \sigma^2_{\epsilon}$.

\subsubsection*{Updating $\mu^{\gamma}_j$}

The full conditional density of $\mu^{\gamma}_j$ is proportional to
\[
f_k\left(a_{\epsilon},\sigma^2_{\epsilon},a_{\gamma},\sigma^2_{\gamma},\beta,\mu^{\epsilon},V^{\epsilon},\mu^{\gamma},V^{\gamma}\right)
\exp\left\{-\frac{{\mu^{\gamma}_j}^2}{(1-a_{\gamma})\sigma^2_{\gamma}}\right\}
\]
and the parameter is updated using the adaptive Metropolis-Hastings random walk step.

\subsubsection*{Updating $V^{\gamma}_j$}

The full conditional density of $V^{\gamma}_j$ is proportional to
\[
f_k\left(a_{\epsilon},\sigma^2_{\epsilon},a_{\gamma},\sigma^2_{\gamma},\beta,\mu^{\epsilon},V^{\epsilon},\mu^{\gamma},V^{\gamma}\right)
\left(1-V^{\gamma}_j\right)^{M_{\gamma}-1}
\]
and the parameter is updated using the adaptive Metropolis-Hastings random walk step after taking the transformation $\log V^{\gamma}_j - \log \left(1- V^{\gamma}_j\right)$.

\subsubsection*{Updating $M_{\gamma}$}

The full conditional distribution of $M_{\gamma}$ is $\Be\left(a^{\star}, b^{\star}\right)$ where $a^{\star}=N_k$ and $b^{\star} = 1 - \sum_{j=1}^{N_k} \log \left(1-V^{\gamma}_j\right)$.

\subsubsection*{Updating $a_{\gamma}$}

The full conditional density of $a_{\gamma}$ is proportional to
\[
f_k\left(a_{\epsilon},\sigma^2_{\epsilon},a_{\gamma},\sigma^2_{\gamma},\beta,\mu^{\epsilon},V^{\epsilon},\mu^{\gamma},V^{\gamma}\right)
\exp\left\{-\frac{\sum_{j=1}^{N_k}\mu^{\gamma\,2}_j}{(1-a_{\gamma})\sigma^2_{\gamma}}\right\}
\left(1-a_{\gamma}\right)^{18-N_k/2}
\]
and the parameter is updated using the adaptive Metropolis-Hastings random walk step after taking the transformation $\log a_{\gamma} - \log \left(1- a_{\gamma}\right)$.

\subsubsection*{Updating $\sigma^2_{\gamma}$}

The full conditional density of $\sigma^2_{\gamma}$ is proportional to
\[
f_k\left(a_{\epsilon},\sigma^2_{\epsilon},a_{\gamma},\sigma^2_{\gamma},\beta,\mu^{\epsilon},V^{\epsilon},\mu^{\gamma},V^{\gamma}\right)
\exp\left\{-\frac{\sum_{j=1}^{N_k}{\mu^{\gamma}_j}^2}{(1-a_{\gamma})\sigma^2_{\gamma}}\right\}
\left(\sigma^2_{\gamma}\right)^{-N_k/2}\left(1+\sigma^2_{\gamma}\right)^{-1}
\]
and the parameter is updated using the adaptive Metropolis-Hastings random walk step after taking the transformation $\log \sigma^2_{\gamma}$.

\subsubsection*{Updating $s^{\epsilon}$}

 The full conditional distribution of $s^{\epsilon}_{it}$
for  $i=1,\dots,n$ and $t=1,\dots,T$
 is
\[
p(s^{\epsilon}_{it}=j)\propto
\exp\left\{-\frac{1}{2}\left[\frac{\sum_{l=1}^T e_{il}^{\epsilon\,2}}{a_{\epsilon}\sigma^2_{\epsilon}}
- \frac{r^{\epsilon\,2}_i}{d}
\right]
\right\},
\qquad j=1,\dots,N_k
\]
where $
e_{il}^{\epsilon}=y_{il} - X_{il}\beta - \mu^{\epsilon}_j + \bar{\mu^{\epsilon}}$ and 
$
r^{\epsilon}_i=
\frac{\sum_{l=1}^T e^{\epsilon}_{il}}{a_{\epsilon}\sigma^2_{\epsilon}}+
\frac{d^{\gamma}_i}{a_{\gamma}\sigma^2_{\gamma}}$

\subsubsection*{Updating $s^{\gamma}$}

The full conditional distribution of $s^{\gamma}_i$ for $i=1,\dots,n$ is 
\[
p\left(s^{\gamma}_i=j\right)\propto
\exp\left\{-\frac{1}{2}\left[ \frac{e_i^{\gamma\, 2}}{a_{\gamma}\sigma^2_{\gamma}}
- \frac{r^{\gamma\,2}_i}{d}
\right]
\right\},\qquad  k=1,\dots,N_k
\]
where $
e_{i}^{\gamma}=\mu^{\gamma}_j - \bar{\mu^{\gamma}}$ and
$r^{\gamma}_i=
\frac{\sum_{t=1}^T d^{\epsilon}_{it}}{a_{\epsilon}\sigma^2_{\epsilon}}+
\frac{e^{\gamma}_i}{a_{\gamma}\sigma^2_{\gamma}}$.

\subsubsection*{Updating $\beta$}

We sample $\gamma_i\sim\N\left(c_i/d,1/d\right)$ for $i=1,\dots,n$ which
 allows us to simulate $\beta$ as
$
\beta\sim\N\left(
\mu^{\star},\Sigma^{\star}
\right)
$
where
\[
\mu^{\star}=
\left(\frac{\sum_{i=1}^n \sum_{t=1}^T X_{it}'X_{it}}{a_{\epsilon}\sigma^2_{\epsilon}}+10^{-6} I_p\right)^{-1}
\left(
\frac{
\sum_{i=1}^n \sum_{t=1}^T\left(
y_{it} - \gamma_i - \mu^{\epsilon}_{s^{\epsilon}_{it}}+\bar{\mu^{\epsilon}}\right)}{a_{\epsilon}\sigma^2_{\epsilon}}
\right)
\]
and
\[
\Sigma^{\star}=\left(\frac{\sum_{i=1}^n \sum_{t=1}^T X_{it}'X_{it}}{a_{\epsilon}\sigma^2_{\epsilon}}+10^{-6} I_p\right)^{-1}.
\]

\subsection{A nonparametric  time series model}

The distribution of $\epsilon_t\vert \epsilon_{t-1}$ is approximated by the $k$-th truncated version
\[
p(\epsilon_t\vert \epsilon_{t-1})=\frac{\sum_{j=1}^{N_k} p^{\epsilon}_j \,\N\left(\left.\left(\begin{array}{c}\epsilon_{t-1}\\ \epsilon_t\end{array}\right)\right\vert \left(\begin{array}{c}\mu^{\epsilon}_j\\\mu^{\epsilon}_j\end{array}\right),a_{\epsilon}\sigma^2_{\epsilon}\left(\begin{array}{cc} 1 & \rho_j\\
\rho_j & 1
\end{array}
\right)
\right)}
{\sum_{j=1}^{N_k} p^{\epsilon}_j \,\N\left(\epsilon_{t-1}\left\vert \mu^{\epsilon}_j, a_{\epsilon}\sigma_{\epsilon}^2\right.\right)}
\]
and distribution of the initial value is
\[
p(\epsilon_1)=\sum_{j=1}^{N_k} p^{\epsilon}_j \,\N\left(\epsilon_1\left\vert \mu^{\epsilon}_j,a_{\epsilon}\sigma_{\epsilon}^2\right.\right).
\]
where $p_j^{\epsilon}=\frac{V^{\epsilon}_j\prod_{l<j}\left(1-V^{\epsilon}_l\right)}{1-\prod_{l=1}^{N_k} (1-V^{\epsilon}_l)}$ with $V^{\epsilon}_j\sim\Be(1,M_{\epsilon})$ for $j=1,\dots,N_k$, which  is the RSB truncation.
 The initial parameters  are $\phi_1=\left(V^{\epsilon}_{1:N_1}, \mu^{\epsilon}_{1:N_1},\rho_{1:N_1}\right)$
and the extra parameters  added to form the $k$-th truncation are 
 $\phi_k=\left(V^{\epsilon}_{N_k}, \mu^{\epsilon}_{N_k},\rho_{N_k}\right)$. 

The nonparametric mixture model for $\nu_t$ is not constrained and so the algorithm uses the P\'olya urn scheme representation 
to sample  the parameters of this part of the model.
 Let $s^{\alpha}_2,\dots,s^{\alpha}_T$ be the allocation variables for $\nu_1,\dots,\nu_T$ respectively, 
$K_{\alpha}$ be the number of distinct values in the sample
$s^{\alpha}_1,\dots,s^{\alpha}_T$ and $\mu^{\alpha}_1,\dots,\mu^{\alpha}_{K_{\alpha}}$ be the distinct values. 
It is useful to define
\begin{align*}
f_k(a_{\epsilon},\sigma^2_{\epsilon},\mu^{\epsilon},p^{\epsilon},\rho,\alpha)=&
\prod_{t=2}^T
\left(
\frac{\sum_{j=1}^{N_k} p_j^{\epsilon}
(1-\rho_j^2)^{-1/2}
\exp\left\{
-\frac{1}{2}\frac{b_{t,j}^2+b_{(t-1),j}^2-2\rho_j b_{t,j}b_{t-1,j}}{a_{\epsilon}\sigma^2_{\epsilon} (1 - \rho_j)^2}
\right\}}
{\sum_{j=1}^{N_k} p^{\epsilon}_j \exp\left\{
-\frac{1}{2}\frac{b_{t-1,j}^2}{a_{\epsilon}\sigma^2_{\epsilon}}
\right\}}
\right)\\
&\times(a_{\epsilon}\sigma^2_{\epsilon})^{T/2}
\sum_{j=1}^{N_k} p_j^{\epsilon}
\exp\left\{
-\frac{1}{2}\frac{b_{1,j}^2}{a_{\epsilon}\sigma^2_{\epsilon}}
\right\}
\end{align*}
where $b_{t,j} = y_t - \alpha_t - \mu^{\epsilon}_j + \bar{\mu^{\epsilon}}$ for $t=1,\dots,T$ and $j=1,\dots,N_k$
with  values of the parameter from the $k$-th truncation.
The MCMC sampler updates many parameters using a variation of the adaptive random walk algorithm of
\cite{atros05} described in section A.1.

\subsubsection*{Updating $\mu^{\epsilon}_j$}

The full conditional density of $\mu^{\epsilon}_j$ is proportional to
\[
f_k\left(a_{\epsilon},\sigma^2_{\epsilon},\mu^{\epsilon},p^{\epsilon},\rho,\alpha\right)
\exp\left\{-\frac{{\mu^{\epsilon}_j}^2}{(1-a_{\epsilon})\sigma^2_{\epsilon}}\right\}
\]
and the parameter is updated using the adaptive Metropolis-Hastings random walk step.

\subsubsection*{Updating $V^{\epsilon}_j$}

The full conditional density of $V^{\epsilon}_j$ is proportional to
\[
f_k\left(a_{\epsilon},\sigma^2_{\epsilon},\mu^{\epsilon},p^{\epsilon},\rho,\alpha\right)
\left(1-V^{\epsilon}_j\right)^{M_{\epsilon}-1}
\]
and the parameter is updated using the adaptive Metropolis-Hastings random walk step after taking the transformation 
$\log V^{\epsilon}_j - \log \left(1- V^{\epsilon}_j\right)$.

\subsubsection*{Updating $\rho_j$}

The full conditional density of $\rho_j$ is proportional to
\[
f_k\left(a_{\epsilon},\sigma^2_{\epsilon},\mu^{\epsilon},p^{\epsilon},\rho,\alpha\right)
\]
and the parameter is updated using the adaptive Metropolis-Hastings random walk step after taking the transformation 
$\log \left(1+\rho_j\right) - \log \left(1- \rho_j\right)$.

\subsubsection*{Updating $M_{\epsilon}$}

The full conditional distribution of $M_{\epsilon}$ is $\Be\left(a^{\star},b^{\star}\right)$ where 
$a^{\star}=1+N_k$ and $b^{\star}=1 - \sum_{j=1}^{N_k} \log \left(1-V^{\epsilon}_j\right)$.

\subsubsection*{Updating $a_{\epsilon}$}

The full conditional density of $a_{\epsilon}$ is proportional to
\[
f_k\left(a_{\epsilon},\sigma^2_{\epsilon},\mu^{\epsilon},p^{\epsilon},\rho,\alpha\right)
\exp\left\{-\frac{\sum_{j=1}^{N_k}{\mu^{\epsilon}_j}^2}{(1-a_{\epsilon})\sigma^2_{\epsilon}}\right\}
(1-a_{\epsilon})^{18-N_k/2}
\]
and the parameter is updated using the adaptive Metropolis-Hastings random walk step after taking the transformation $\log a_{\gamma} - \log \left(1- a_{\gamma}\right)$.

\subsubsection*{Updating $\sigma^2_{\epsilon}$}

The full conditional density of $\sigma^2_{\epsilon}$ is proportional to
\[
f_k\left(a_{\epsilon},\sigma^2_{\epsilon},\mu^{\epsilon},p^{\epsilon},\rho,\alpha\right)
\exp\left\{-\frac{\sum_{j=1}^{N_k}{\mu^{\epsilon}_j}^2}{(1-a_{\epsilon})\sigma^2_{\epsilon}}\right\}
\left(\sigma^2_{\epsilon}\right)^{-N_k/2}\left(1+\sigma^2_{\epsilon}\right)^{-1}
\]
and the parameter is updated using the adaptive Metropolis-Hastings random walk step after taking the transformation $\log \sigma^2_{\epsilon}$.



\subsubsection*{Updating $\alpha_1,\dots,\alpha_T$}

The full conditional density of $\alpha_1$ is proportional to
\[
\sum_{j=1}^{N_k} p_j^{\epsilon}
(1-\rho_j^2)^{-1/2}
\exp\left\{
-\frac{1}{2}\frac{b_{2,j}^2+b_{1,j}^2-2\rho_j b_{2,j}b_{1,j}}{a_{\epsilon}\sigma^2_{\epsilon} (1 - \rho_j)^2}
\right\}\exp\left\{-\frac{1}{2}\frac{\alpha_1^2}{\sigma^2_0}\right\}.
\]
The full conditional density of $\alpha_t$ for $t=2,\dots,T-1$ is proportional to
\[
\prod_{k=t}^{t+1}
\frac{\sum_{j=1}^{N_k} p_j^{\epsilon}
(1-\rho_j^2)^{-1/2}
\exp\left\{
-\frac{1}{2}\frac{b_{k,j}^2+b_{k-1,j}^2-2\rho_j b_{k,j}b_{k-1,j}}{a_{\epsilon}\sigma^2_{\epsilon} (1 - \rho_j)^2}
\right\}}
{\sum_{j=1}^{N_k} p_j^{\epsilon}\exp\left\{
-\frac{1}{2}\frac{b_{k-1,j}^2}{a_{\epsilon}\sigma^2_{\epsilon}}
\right\}}
\exp\left\{-\frac{1}{2}\sum_{k=t}^{t+1}\frac{\left(\alpha_k-\alpha_{k-1}-\mu^{\alpha}_{s^{\alpha}_k}\right)^2}{\sigma^2_{\alpha}a_{\alpha}}\right\}.
\]
The full conditional density of $\alpha_T$ is proportional to
\[
\frac{\sum_{j=1}^{N_k} p_j^{\epsilon}
(1-\rho_j^2)^{-1/2}
\exp\left\{
-\frac{1}{2}\frac{b_{T,j}^2+b_{T-1,j}^2-2\rho_j b_{T,j}b_{T-1,j}}{a_{\epsilon}\sigma^2_{\epsilon} (1 - \rho_j)^2}
\right\}}
{\sum_{j=1}^{N_k} p_j^{\epsilon}\exp\left\{
-\frac{1}{2}\frac{b_{T-1,j}^2}{a_{\epsilon}\sigma^2_{\epsilon}}
\right\}}
\exp\left\{-\frac{1}{2}\frac{\left(\alpha_T-\alpha_{T-1}-\mu^{\alpha}_{s^{\alpha}_T}\right)^2}{\sigma^2_{\alpha}a_{\alpha}}\right\}
\]


\subsubsection*{Updating $s^{\alpha}_t$}

The full conditional distribution of $s^{\alpha}_t$ is
\[
p(s^{\alpha}_t=j)\propto \left(\sum_{l=2; l\neq t}^T \I(s^{\alpha}_l=j)\right)
a_{\alpha}^{-1/2}
\exp\left\{-\frac{1}{2}\frac{\left(\alpha_t-\alpha_{t-1}-\tilde\mu_j^{\alpha}\right)^2}{a_{\alpha}\sigma^2_{\alpha}}\right\}
,\qquad j=1,\dots,K^{-}_{\alpha}
\]
and
\[
p(s^{\alpha}_t=K^{-}_{\alpha}+1)\propto M_{\alpha}
\exp\left\{-\frac{1}{2}\frac{\left(\alpha_t-\alpha_{t-1}\right)^2}{\sigma^2_{\alpha}}\right\}
\]
where $K^{-}_{\alpha}$ is the number of distinct values without the $t$-th allocation and 
$\tilde\mu_1^{\alpha},\dots,\tilde\mu_{K^{-}_{\alpha}}^{\alpha}$ are the corresponding distinct values.

\subsubsection*{Updating $\mu^{\alpha}_j$}

The full conditional distribution of $\mu^{\alpha_j}$ is $\N\left(\mu^{\star},\sigma^{\star\,2}\right)$ where
\[
\mu^{\star}=\frac{\sum_{t=2}^T \I(s_t^{\alpha}=j)\left(\alpha_t-\alpha_{t-1}\right)/a_{\alpha}}{\sum_{t=2}^T \I(s_t^{\alpha}=j)/a_{\alpha}+1/(1-a_{\alpha})}
\]
and
\[
\sigma^{\star\,2}=\frac{\sigma^2_{\alpha}}{\sum_{t=2}^T \I(s_t^{\alpha}=j)/a_{\alpha}+1/(1-a_{\alpha})}.
\]

\subsubsection*{Updating $a_{\alpha}$}

The full conditional density of $a_{\alpha}$ is proportional to
\[
{a_{\alpha}}^{-(T-1)/2}
\left(1-a_{\alpha}\right)^{18-K_{\alpha}/2}
 \exp\left\{-\frac{1}{2}
\sum_{t=2}^T
\frac{\left(\alpha_t-\alpha_{t-1}-\mu^{\alpha}_{s^{\alpha}_i}\right)^2}{a_{\alpha}\sigma^2_{\alpha}}\right\}
\exp\left\{-\frac{\sum_{j=1}^{K_{\alpha}}\mu^{\alpha\,2}_j}{(1-a_{\alpha})\sigma^2_{\alpha}}\right\}
\]
and the parameter is updated using the adaptive Metropolis-Hastings random walk step after taking the transformation $\log a_{\alpha} - \log \left(1- a_{\alpha}\right)$.

\subsubsection*{Updating $\sigma^2_{\alpha}$}

The full conditional density of $\sigma^2_{\alpha}$ is proportional to
\[
{\sigma^2_{\alpha}}^{-(T-1+K_{\alpha})/2}
 \exp\left\{-\frac{1}{2}
\sum_{t=2}^T
\frac{\left(\alpha_t-\alpha_{t-1}-\mu^{\alpha}_{s^{\alpha}_i}\right)^2}{a_{\alpha}\sigma^2_{\alpha}}\right\}
\exp\left\{-\frac{\sum_{j=1}^{K_{\alpha}}\mu^{\alpha\,2}_j}{(1-a_{\alpha})\sigma^2_{\alpha}}\right\}
\left(1+100\sigma^2_{\alpha}\right)^{-1}
\]
and the parameter is updated using the adaptive Metropolis-Hastings random walk step after taking the transformation $\log \sigma^2_{\alpha}$.

\subsubsection*{Updating $M_{\alpha}$}

The full conditional density of $M_{\alpha}$ is proportional to
\[
\frac{\Gamma(M_{\alpha})}{\Gamma(M_{\alpha}+T-1)}M_{\alpha}^{K_{\alpha}}
\]
and the parameter is updated using the adaptive Metropolis-Hastings random walk step after taking the transformation $\log M_{\alpha}$.

\end{document}